\documentclass[useAMS,usenatbib]{mnras}
\usepackage{amsmath,psfig,epsfig,graphics}
\usepackage{graphicx}
\usepackage{epstopdf}
\usepackage{float}
\usepackage{enumerate}
\usepackage{natbib}
\usepackage{threeparttable}
\usepackage{longtable}
\usepackage[normalem]{ulem}

\def\chandra    {{\em Chandra}\/}

\def\xmm        {XMM-{\em Newton}\/}

\def\vla        {{\em VLA}\/}
\def\gmrt       {{\em GMRT}\/}

\def\arcmin{\hbox{$^\prime$}}
\def\arcsec{\hbox{$^{\prime\prime}$}}

\usepackage{tikz,xcolor,hyperref}
\definecolor{lime}{HTML}{A6CE39}
\DeclareRobustCommand{\orcidicon}{%
    \begin{tikzpicture}
    \draw[lime, fill=lime] (0,0) 
    circle [radius=0.16] 
    node[white] {{\fontfamily{qag}\selectfont \tiny ID}};   \draw[white, fill=white] (-0.0625,0.095) 
    circle [radius=0.007];  \end{tikzpicture}
    \hspace{-2mm}}
\foreach \x in {A, ..., Z}{%
    \expandafter\xdef\csname orcid\x\endcsname{\noexpand\href{https://orcid.org/\csname orcidauthor\x\endcsname}{\noexpand\orcidicon}}
    }

\newcommand{\RNum}[1]{\uppercase\expandafter{\romannumeral #1\relax}}

\begin{document}

\title[X-ray Cool Core Remnants]{X-ray Cool Core Remnants Heated by Strong Radio AGN Feedback}

\author[Liu et al.]
{Wenhao Liu${}^{1}$\thanks{E-mail: whliu@pmo.ac.cn}\orcidA{}, 
Ming Sun${}^2$\thanks{E-mail: ms0071@uah.edu}\orcidB{},
G. Mark Voit${}^3$\orcidC{},
Dharam Vir Lal${}^4$\orcidD{},
Paul Nulsen${}^{5,6}$\orcidE{}, \newauthor
Massimo Gaspari${}^{7}$\orcidF{},
Craig Sarazin${}^8$\orcidG{},
Steven Ehlert${}^{9}$\orcidH{},
Xianzhong Zheng${}^{1}$\orcidI{}
\newauthor 
\\
$^{1}$ Purple Mountain Observatory, Chinese Academy of Science, 10 Yuanhua Road, Nanjing 210023, China \\
$^{2}$ Department of Physics and Astronomy, University of Alabama in Huntsville, Huntsville, AL 35899, USA\\
$^{3}$ Department of Physics and Astronomy, Michigan State University, East Lansing, MI 48824, USA \\
$^{4}$ National Centre for Radio Astrophysics, Pune University Campus, Ganeshkhind, Pune 411 007, India\\
$^{5}$ Harvard-Smithsonian Center for Astrophysics, 60 Garden Street, Cambridge, MA 02138, USA \\
$^{6}$ ICRAR, University of Western Australia, 35 Stirling Hwy, Crawley, WA 6009, Australia\\
$^{7}$ Department of Astrophysical Sciences, Princeton University, 4 Ivy Lane, Princeton, NJ 08544-1001, USA \\
$^{8}$ Department of Astronomy, University of Virginia, Charlottesville, VA 22904, USA\\
$^{9}$NASA Marshall Space Flight Center, Huntsville, AL 35812, USA \\
}


\maketitle

\label{firstpage}

\begin{abstract}
	Strong AGN heating provides an alternative means for the disruption of cluster cool cores (CCs) to cluster mergers. 
	In this work we present a systematic \chandra\ study of a sample of 108 nearby ($z<0.1$) galaxy clusters, 
	to investigate the effect of AGN heating on CCs.
	About 40\% of clusters with small offsets between the BCG and the X-ray centre ($\le50$ kpc) have small CCs. 
	For comparison, 14 of 17 clusters with large offsets have small CCs, which suggests that mergers or sloshing can be efficient in reducing the CC size.
	Relaxed, small CC clusters generally have weak radio AGNs ($P_{1.4\rm GHz}<10^{23}$ W~Hz$^{-1}$), and they 
	show a lack of systems hosting a radio AGN with intermediate radio power ($2\times10^{23}<P_{1.4\rm GHz}<2\times10^{24}$ W~Hz$^{-1}$).
	We found that the strongest circumnuclear ($<1$ kpc) X-ray emission only exists in clusters with strong radio AGN.
	The duty cycle of relaxed, small CC clusters is less than half of that for large CC clusters. It suggests that the radio activity of BCGs is affected by the properties of the surrounding gas beyond the central $\sim10$ kpc, and strong radio AGNs in small X-ray CCs
	fade more rapidly than those embedded in large X-ray CCs. A scenario is also presented for the transition of large CCs and coronae due to radio AGN feedback.
	We also present a detailed analysis of galaxy cluster 3C~129.1 as an example of a CC remnant possibly disrupted by radio AGN.
\end{abstract}

\section{Introduction}
\noindent
As the most massive virialized structure in the Universe, clusters of galaxies are excellent laboratories to study 
physical processes in structure formation.
Galaxy clusters are permeated with the hot, X-ray emitting intra-cluster medium (ICM). When the radiative cooling time of the hot gas is shorter than the age of the system, in the absence of heating, the hot gas cools, condenses, and flows toward the centre \citep[e.g.,][]{Fabian94}.
However, much less cool gas has been found by \chandra\ and \xmm\ than the prediction from the cooling flow models 
\citep[e.g.,][]{David01, Peterson01}, suggesting there is a heating source compensating for the radiative cooling.
Now, feedback from the central active galactic nuclei (AGN) appears to be the most promising heating source among
many possibilities, and it has been widely accepted that AGN feedback plays an important role in galaxy formation and evolution
\citep[e.g.,][]{McNamara07,Fabian12,Gaspari2020,Eckert2021,Hlavacek2022, Donahue2022,Heckman2023}.

AGN outflows may simultaneously explain the quenching of
star formation in massive galaxies, the exponential cut-off at the bright end of the galaxy luminosity function, 
the supermassive black hole mass-bulge mass relation, and the quenching of cooling-flows in cluster cores 
\citep[e.g.,][]{Scannapieco05,Begelman05,Croton06}.
Outflows from radio AGN are especially important locally because nearly
all bright radio galaxies are hosted by early-type galaxies that dominate the high end of the luminosity function
and the galaxy population in clusters. 
Radio-loud AGNs favor dense environments and many are found in galaxy groups and
clusters, embedded in the X-ray emitting intracluster medium (ICM) and intragroup medium (IGM) \citep[e.g.,][]{Best05}.
In the nearby Universe, a radio-loud AGN, usually hosted by the brightest cluster galaxy (BCG) of a cluster
or a group, can drive energetic radio jets, which extend outwards, inflate radio lobes, and create cavities and/or shocks 
visible in X-ray images \citep[e.g.,][]{Churazov2001}.
The great mechanical power of radio AGN can not only quench 
cooling in cluster cool cores, but also drive the ICM properties away from those defined by simple self-similar relations
involving only gravity \citep[e.g.,][]{McNamara07,Sun12}.

With the help of the superior angular resolution of \chandra, the relation between radio AGN and X-ray cool cores has been established.
Cluster cores seem to be bimodal in the core entropy distribution, which is naturally connected with cool core (CC) and non-cool core (NCC)
clusters respectively \citep[e.g.,][]{Cavagnolo09}. This CC/NCC classification can also be determined using central cooling time, central temperature
gradient, central density, surface brightness cuspiness, and central 
entropy \citep[e.g.,][]{Leccardi2008, Hudson2010, McDonald2013, McDonald2017,Ghirardini2019}. 
However, intermediate objects may be classified differently using different diagnostic methods.
While radio AGNs are typically enhanced in CC clusters, many radio AGNs are present in NCC clusters. 
\citet{Sun09b} showed that the traditional CC/NCC dichotomy is too simple and the CCs hosting BCG with a strong radio AGN ($P_{\rm 1.4 GHz}>2\times10^{23}$ W~Hz$^{-1}$) can be divided into two classes: 
the large cool core (LCC) class and the corona class based on the $0.5-2.0$ keV luminosity within the cool core.
Small coronae, which can be misidentified as X-ray AGNs at $z>0.1$, are mini-cool cores in groups and clusters.
Thus, the distinction between CCs and NCCs is more complicated, as some NCC clusters actually have small coronae in their centres.
The corona fraction is at least comparable with that of LCCs for clusters hosting BCGs with strong radio AGNs in the \citet{Sun09b} sample and
some coronae may be the remnants of LCC \citep[e.g.,][]{Sun09b}.

The CC/NCC distribution is assumed to result from different evolutionary histories of galaxy clusters, but the origin of this distribution is 
not well understood. The transition between CC clusters and NCC clusters is important to understand the formation and evolution 
of cluster CCs \citep[see][]{Burns08}. A joint analysis of \chandra\ X-ray and South Pole Telescope Sunyaev-Zel'dovich 
observations found that the CC
fraction in a sample of clusters at $0.3<z<1.3$ is remarkably stable although the mass of clusters has increased by a factor of 
$\sim4$ over the past 9 Gyr \citep{Ruppin2021}. This suggests that the CC disruption is balanced well by CC restoration 
(e.g., a NCC cluster relaxes to the CC state) since high redshift \citep{McDonald2017}.
Using a multi-temperature spectral analysis, \citet{Molendi2023} 
found that there may exist a small amount of low entropy gas with short cooling times in the subclass of NCC clusters
known as CC remnant systems, therefore allowing a transformation to CC systems on a relatively short timescale.
One of the main mechanisms to transform CC to NCC clusters is cluster mergers. Using a well-defined X-ray selected cluster sample,
\citet{Rossetti2011} found that none of the dynamically disturbed systems, often showing large-scale radio halos,
can be classified as a CC \citep[e.g.,][]{Cassano2013, Cuciti2015}.
Simulations suggest that CCs can only 
be destroyed by high-energy, low angular momentum major mergers \citep[e.g.,][]{Hahn2017}. A recent study of galaxy cluster 
A1763 shows that large-scale gas-sloshing could also disrupt cluster CCs and establish NCC \citep{Douglass2018}.
However, the physics related to core restoration and destruction is very difficult to capture in simulations.

The alternative mechanism for CC disruption is by powerful outbursts from the central AGN
through the cold AGN feedback mechanism. That is, the hot gas in the cores of galaxy groups/clusters
cools to form a multiphase medium consisting of cold clouds that 
fall inward and feed the supermassive black hole (SMBH). The accretion rate can be boosted by orders of magnitude 
via a process known as chaotic cold accretion (CCA)/cold gas 
precipitation \citep[e.g.,][]{Pizzolato2005,Sharma12,Gaspari2013,Voit15b,Voit2019,Qiu2021,Mikinley2022,Wittor2023}.
The large amount of energy released by an energetic AGN outburst can disrupt the CCs and leave CC remnants.
They may be more common in groups, where CCs are smaller than those in clusters. The impact of AGN outbursts is much more pronounced in 
low-mass systems due to their shallow gravitational potential \citep[e.g.,][]{Giodini10, Sun12}. X-ray cavities have been detected in only a few 
galaxy groups hosting radio luminous AGNs ($P_{\rm 1.4 GHz}>10^{24}$ W Hz$^{-1}$), e.g., NGC~4782 \citep{Machacek2007},  NGC~4261 \citep{OSullivan2017},
PKS~B2152-699 \citep{Worrall12}, 3C~88 \citep{Liu_3C88}, IC~4296 \citep{Grossova2019}.
Fewer cases of X-ray cavities have been reported in groups than in clusters due to the relatively low X-ray luminosity of the smaller 
X-ray atmospheres that enclose the lobes of luminous radio AGN in groups.
In another way, CCs could have been disrupted by AGN 
outbursts and turned into coronae in galaxy groups \citep[e.g.,][]{Sun09b,OSullivan2010}. CC remnants also exist in galaxy clusters. \citet{Rossetti2010} studied the cores of a sample
of 35 clusters with \xmm\ data and selected 12 CC remnants based on the excess metal abundances near the cluster centres. These systems generally
have rather dense cores and most of them may be classified as weak CCs \citep[e.g.,][]{Hudson2010}. 
One of them (A3558) also hosts a luminous, small corona.

While the CCs in galaxy groups can be more easily disrupted by AGN outburst, in this paper, we aim to use a sample of 
nearby hot galaxy clusters to study how transitions between the states of cluster cores are related to radio AGN feedback.
The paper is organized as follows. The cluster sample selection is defined in Section \ref{Sample}. 
The data analysis is presented in Section \ref{DataAnalysis}. In Section \ref{Results} we present the results.
After studying the general properties of the sample, we present a detailed analysis of the cluster 3C~129.1 with a CC remnant, associated
with a strong radio AGN in section \ref{Case}. Discussions are in Section \ref{Discussion} and Section \ref{Conclusion} contains the
conclusions. Throughout the paper, we assume a cosmology with $H_{0}=70$ km s$^{-1}$ Mpc$^{-1}$, $\Omega_{M}=0.3$, and $\Omega_{\Lambda}=0.7$.

\section{Sample Selection}
\label{Sample}
We want to examine the relationship between the radio AGN in the central BCGs and the transitions between X-ray LCCs and coronae
of nearby hot galaxy clusters. 
The corona, typically with a low X-ray luminosity, can be easily missed at $z>0.1$.
In order to study the properties of cluster CCs, especially for clusters with small coronae,
we need to resolve the properties of hot gas (e.g., temperature and density) to the very central region, e.g., within a few kpc.
Therefore, we selected our sample from the Meta-Catalogue of X-ray detected Clusters of galaxies \citep[MCXC,][]{Piffaretti2011}
based on the following criteria: (1) $z<0.1$; (2) $kT_{\rm ICM}>3$ keV (The temperatures were obtained 
from the previous studies in the literature or from our own analysis). Here we chose a temperature cut of 3 keV since we focused on 
hot galaxy clusters instead of galaxy groups;
(3) with \chandra\ observations (due to its unprecedented spatial resolution); (4) $>\sim35$ net counts in the 0.5-3.0 keV band within the central 
5 kpc in order to constrain the properties of the coronae if exist. 
The constraints on the CC properties depend on many factors, such as the \chandra\ exposure, the distance of cluster, the Galactic 
absorption and so on. Therefore, we first used this low threshold of 35 counts to include as many clusters as we could. We then examined the 
clusters case by case and excluded those clusters where we could not make statistical constraints on the CC properties.
Finally, we have a total of 108 galaxy clusters in our sample.

\begin{figure}
	\hbox{\hspace{-5px}
\includegraphics[scale=0.42]{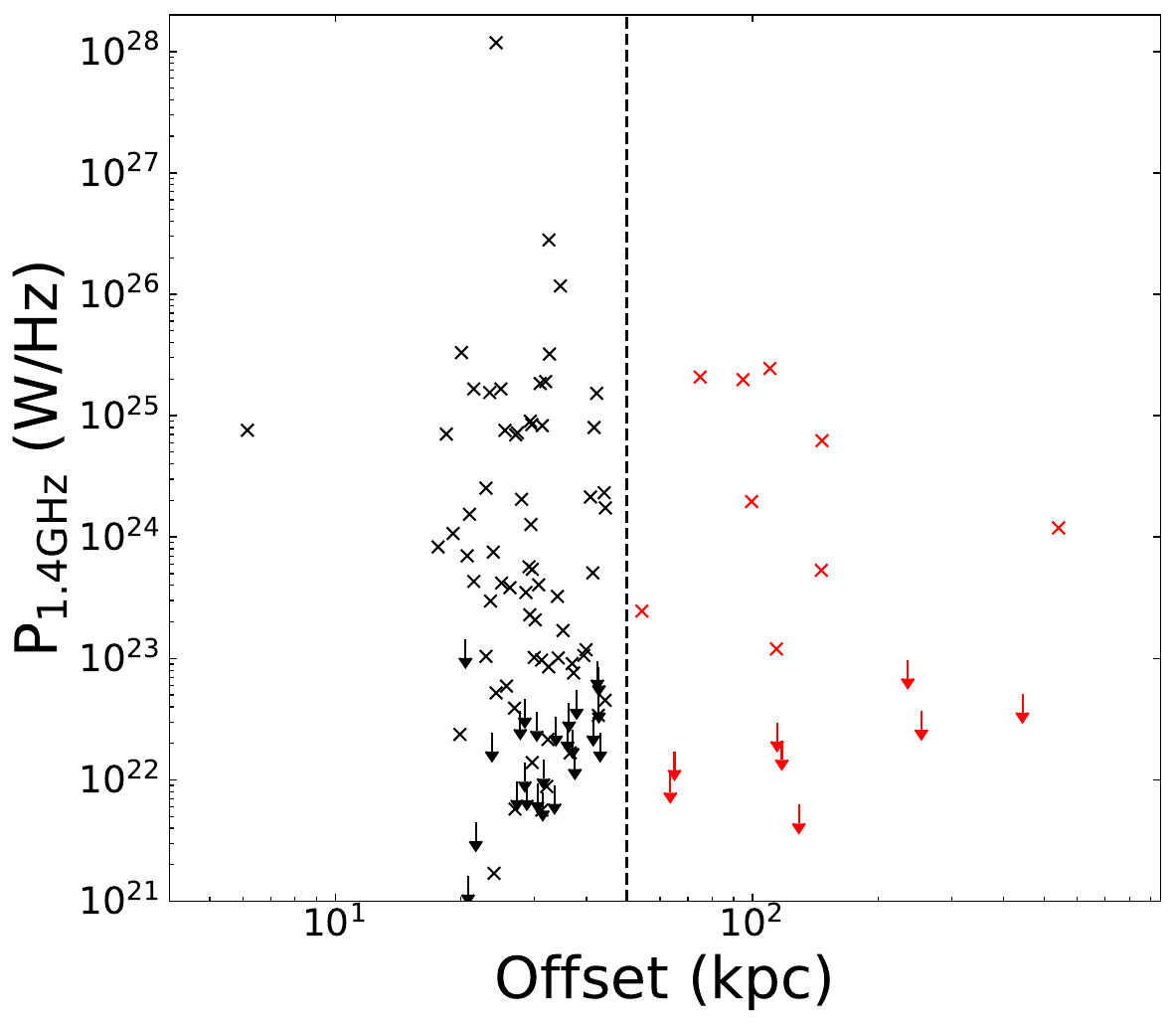}
}
\caption{
	The radio luminosity of central BCGs at 1.4 GHz vs. the offsets between the positions of the BCG and the X-ray cluster centre.
	Clusters with the offsets greater than 50 kpc are shown in red. 
}
\label{offset}
\end{figure}

Our main interest is to investigate the effect of radio AGN heating by the central BCGs on the CCs of galaxy clusters.
However, some clusters in our sample show signatures of recent mergers, e.g., large offsets between
the central BCG, the most luminous galaxy in the Two Micron All Sky Survey (2MASS) $K_{\rm s}$ band in a cluster, 
and the X-ray cluster centre.
The X-ray centre for each cluster is determined with the SPA code\footnote{https://sites.google.com/site/adambmantz/work/morph14} by computing the median photon location in an iteratively shrinking aperture, and this median centre
can compromise between the X-ray peak and the cluster centroid \citep{Mantz2015}. 
In this study, we used the offset as the separation between the BCG and this median centre instead of
the separation between the BCG and the X-ray peak.
In Fig. \ref{offset}, we plot the offsets between the central BCG (its position is obtained from 2MASS K-band image) 
and X-ray cluster centre vs. the radio luminosities of the central BCGs at 1.4 GHz.
There are 17 systems with offsets greater than 50 kpc (if we change the threshold to 60 kpc, the number of clusters
with offsets greater than the threshold decrease only by one).
The radio fluxes of the central BCGs are obtained from the NRAO VLA Sky Survey (NVSS) and the Faint 
Images of the Radio Sky at Twenty Centimeter (FIRST) survey. If neither of them is available, 
we used radio surveys in other bands, e.g., the Sydney University Molonglo Sky Survey (SUMSS) at 843 MHz, 
and converted to the flux at 1.4 GHz assuming a radio spectral index of $\alpha=0.8$ (defined as $S\propto \nu^{-\alpha}$ where
$S$ is the radio flux and $\nu$ is the frequency).
In our sample $\sim55$\% of BCGs (59 out of 108) have 
$P_{\rm 1.4GHz} > 10^{23}$ W Hz$^{-1}$, and $\sim31$\% of BCGs (34 out of 108) have
$P_{\rm 1.4GHz} > 10^{24}$ W Hz$^{-1}$. 
In Table \ref{tab1} we list the properties of our cluster sample. 

\begin{table}
\caption{Cluster Sample Properties}
\begin{center}
\begin{tabular}{cc|cc|cccc}
    \hline
	\multicolumn{2}{c|}{$P_{\rm 1.4GHz}^{a}$ (W Hz$^{-1}$)} & \multicolumn{2}{c|}{Offset$^{b}$ (kpc)} & \multicolumn{3}{c|}{$r_{\rm 4 Gyr}^{c}$ (kpc)}\\
	 $>10^{23}$ & $<10^{23}$  & $<50$  &  $>50$ & $>15$  & $<6$ & $6-15$ \\
    \hline
	 59 &  49  &   91  & 17 &   58  &  20    & 30 \\  %
    \hline
\end{tabular}
\end{center}
\label{tab1}
\begin{tablenotes}
\item
 $a$: radio luminosity of central BCG at 1.4 GHz;
	$b$: The offset between the BCG and the X-ray cluster centre.
 $c$: Cooling radius where the gas cooling time is 4 Gyr.
\end{tablenotes}
\end{table}

\section{Data Analysis}
\label{DataAnalysis}
We performed the standard \chandra\ data analysis using CIAO 4.12 and CALDB version 4.9.0. For each observation,
we reprocessed the data from level 1 event files using the {\tt CHANDRA\_REPRO} script to account for afterglow,
bad pixels, charge transfer inefficiency, and time-dependent gain correction.
The improved background filtering was applied for observation taken in Very
Faint (VFAINT) mode by setting {\tt CHECK\_VF\_PHA} as ``yes'' to remove bad events that are likely associated
with cosmic rays. We generated the light curve to remove any period 
affected by background flares. The point sources were detected in the 0.5-7.0 keV count images using 
the CIAO tool {\tt WAVDETECT} and then removed from our data analysis. 
We used the CALDB blank sky background files in our data analysis. For each observation the standard blank sky file for each
chip was reprojected to match the time-dependent aspect solution, and normalized to match the count rate in the
9.5-12.0 keV band to account for variations in the particle background. 
We estimated the X-ray sky background residuals using background models including two APEC components 
with zero redshift and Solar abundance and one power-law component.
For clusters with regions free of cluster emission, we extracted the spectrum from the region where the surface brightness is 
approximately constant. If cluster emission is full of the field of view, we extracted the spectra from two regions, one background-dominated outer region, and one 
source-dominated inner region. We fit two spectra simultaneously, with the normalizations of the background models linked \citep[e.g.,][]{Sun09}.
The best-fit background residual models were then used in spectral and surface brightness analysis.
In fact, the variations of X-ray background have little impact on our results since our study is focused on the bright X-ray cool cores.
For the spectral fitting, we used XSPEC version 12.10.0 and AtomDB 3.0.9 with the solar abundance table of
\citet{aspl}. The absorption model is {\tt TBABS}.

\subsection{Cool Core Size and Luminosity}
\label{SizeLum}
In this study, we aim to study the properties of cluster CCs, and the transitions between the large CCs
and small CCs due to the radio AGN feedback. For this purpose, we need to define the sizes of CCs and estimate their X-ray 
luminosities.

X-ray CCs of galaxy clusters are usually defined based on the cooling time of the gas. 
We used a cooling radius of $r_{\rm 4Gyr}$, where the gas cooling time is 4 Gyr instead of the conventionally
used value of 7.7 Gyr, to quantify the size of CCs. 
Previous study has shown that $r_{\rm 4Gyr}$ can effectively represent the whole region for small coronae \citep{Sun2007},
and further reasons to choose this radius were discussed in \citet{Sun09b}. 
In this work we adopted this cooling radius to distinguish between the LCCs and small coronae. 
A cooling radius with a somewhat different cooling time, e.g., 3-5 Gyr, does not alter
our main conclusions.

To derive the cooling time profiles toward the centres of clusters, 
we generally followed the method used in \citet{Cavagnolo09} to obtain the gas temperature
and density profiles.
First, we extracted the spectra in a series of annuli within the central $\sim100$ kpc centred on the BCG
and explored the properties of the hot gas as close to the centre as possible. 
We then obtained the gas temperature and abundance profiles by fitting an absorbed thermal model to the spectra,
with the fixed background models scaled to the extraction area.
Second, we extracted the surface brightness profiles in the $0.5-7.0$ keV band using concentric annular bins
centred on the BCG.
We deprojected the surface brightness profiles using
the “onion-peeling” method to obtain the gas density profile \citep[e.g.,][]{Kriss1983,Cavagnolo09}, taking into account
the variations of temperature and metal abundance.
We checked if the surface brightness profiles in the soft band (e.g., $0.5-2.0$ keV) are used, the derived density profiles are
consistent with the results obtained using the surface brightness in the $0.5-7.0$ keV band.
Once we had the temperature and density profiles, we calculated the cooling time profiles
using equation $t_{cool}=\frac{3nkT}{2n_{e}n_{H}\Lambda(T,Z)}$, where $n$ is the total number density, $n_{e}$ and $n_{H}$ are the electron
and proton densities, $\Lambda(T,Z)$ is the cooling function for a given temperature and metal abundance.
Finally we calculated the cooling radius $r_{\rm 4Gyr}$ based on the cooling time profile.
In our sample, we defined X-ray CCs as LCCs if $r_{\rm 4Gyr}>15$ kpc, and small CCs if $r_{\rm 4Gyr}\le15$ kpc
(corona class with $r_{\rm 4Gyr}\le6$ kpc and transition CCs if $6<r_{\rm 4Gyr}\le15$ kpc).

Many clusters in our sample were studied by \citet{Cavagnolo09} in detail. They provided entropy profiles that can be
used to distinguish CC and NCC clusters, as well as the cooling time profiles, which can be used
to calculate the cooling radius $r_{\rm 4Gyr}$. However, we noticed that the radial profiles were centered on the ICM X-ray peak in \citet{Cavagnolo09}, while ours were centered on the BCG. Therefore, we used the cooling time 
profiles in \citet{Cavagnolo09} if the offsets between the BCGs and
the cluster centres used in their studies are small than $\sim5$ kpc or $\sim2$ kpc for clusters with LCCs 
or small CCs, respectively.
In Table \ref{tab2}, \ref{tab3} and \ref{tab4}, we indicated which cluster was from our data 
analysis and for which cluster we used their results.

For corona systems, there is usually an enhancement of the surface brightness in the centre \citep[e.g.,][]{Sun2007}.
In our study, if the derived cooling radius $r_{\rm 4Gyr}$ is less than 6 kpc and the surface brightness of the system 
is flat in the centre, we took the cooling radius to be the upper limit for the corona size.
Some clusters do not have either a large CC or a central corona, e.g., A2147 \citep{Sun2007}. 
The central cooling time of these clusters is much larger than 4 Gyr, and they show no central enhancement of X-ray surface brightness.
The previous studies have shown that the X-ray coronae of galaxies in galaxy clusters are quite compact with an extension of
a few kpc, and can extend to $\sim$10 kpc in some cases \citep[e.g.,][]{Sun2007}. Here we used 3 kpc as an upper limit of the CC size
for these galaxy clusters.

Once the CC size was derived, we were able to measure the cooling luminosity. We extracted the spectrum from the cool core region
centred on the BCG within the radius of $r_{\rm 4Gyr}$. The background spectrum
was extracted within the region from $r_{\rm 4Gyr}$ to 2$r_{\rm 4Gyr}$. 
We then fitted the spectrum with an absorbed thermal component
for the thermal gas, and a power-law component to account for any possible emission from the central AGN. 
Unlike in large CC systems, the gas abundance in small CC systems usually cannot be constrained, so we fixed it at 0.8 Solar, 
a typical value found in corona systems \citep[e.g.,][]{Sun2007}. 
For some clusters (e.g., NCC cluster A2147), the extracted spectrum has very limited counts, so we also fixed the temperature 
of the thermal component to 0.7 keV in order to derive an upper limit for the luminosity, as in \citet{Sun2007}.
We measured the bolometric luminosity within the CC in our study.
The properties of each cluster in our sample are summarized in Table \ref{tab2} for clusters with small offsets and large CCs,
Table \ref{tab3} for clusters with small offsets and small CCs, and Table \ref{tab4} for clusters with large offsets.

\begin{table*}
	\protect\caption{Summary of results for clusters with small offsets ($\le$50 kpc) and large cool cores ($r_{\rm 4 Gyr} > 15$ kpc)} 
\scriptsize
\begin{tabular}{|l|c|c|c|c|c|c|c|c|c|}
\hline 
\hline
		Cluster & $z^{a}$ & $P_{\rm 1.4GHz}^{b}$ & offset$^{c}$ & $r_{\rm 4Gyr}^{d}$ & $L_{\rm X,bol}^{e}$ & Note$^{f}$ \\
		&          & log($P_{\rm 1.4GHz}$/W Hz$^{-1}$) &  (kpc) & (kpc)  & ($10^{44}$ erg/s)   \tabularnewline
\hline 
	      2A0335+096 &  0.0346 &  23.0 &   23.0 & $78.0\pm0.7$ &  $1.42\pm0.03$           & 0 \\
		A85      &  0.0557 &  23.6 &   30.7 & $56.6\pm1.7$ &  $1.04_{-0.08}^{+0.01}$  & 1 \\
		A133     &  0.0569 &  24.8 &   18.4 & $52.4\pm1.8$ &  $0.37_{-0.04}^{+0.03}$  & 1 \\
		A426     &  0.0176 &  25.2 &   24.9 & $82.9\pm2.3$ &  $4.00_{-0.20}^{+0.10}$  & 1 \\
		A478     &  0.0860 &  23.8 &   20.7 & $92.8\pm5.6$ &  $7.65_{-0.16}^{+0.03}$  & 1 \\
		A496     &  0.0328 &  23.5 &   23.5 & $55.5\pm5.8$ &  $0.63\pm0.02$           & 1 \\
		A780     &  0.0549 &  26.4 &   32.5 & $59.7\pm1.3$ &  $1.14\pm0.05$           & 1 \\
		A1644    &  0.0475 &  23.7 &   41.4 & $27.0\pm1.9$ &  $0.053\pm0.004$         & 1 \\
		A1650    &  0.0846 &$<$22.6&   36.3 & $43.5\pm5.6$ &  $0.46_{-0.03}^{+0.01}$  & 1 \\
		A1651    &  0.0853 &  23.0 &   39.4 & $24.4\pm11.8$ & $0.083\pm0.012$  & 0 \\
		A1668    &  0.0635 &  23.9 &   23.9 & $49.8\pm3.1$ &  $0.25_{-0.09}^{+0.10}$  & 0 \\
		A1775    &  0.0757 &  23.5 &   34.0 & $39.8\pm1.0$ &  $0.087\pm0.006$         & 0 \\
		A1795    &  0.0633 &  24.9 &   29.6 & $80.5\pm1.7$ &  $3.16_{-0.14}^{+0.13}$  & 0 \\
		A2029    &  0.0779 &  24.9 &    6.2 & $79.9\pm4.0$ &  $4.59_{-0.17}^{+0.15}$  & 1 \\
		A2033    &  0.0809 &  24.9 &   29.3 & $25.5\pm5.7$ &  $0.061\pm{0.006}$       & 0 \\
		A2052    &  0.0345 &  25.2 &   23.4 & $44.8\pm0.6$ &  $0.30_{-0.02}^{+0.01}$  & 1 \\
		A2063    &  0.0341 &  22.6 &   26.9 & $35.8\pm2.2$ &  $0.075_{-0.025}^{+0.019}$       & 0 \\
		A2065    &  0.0729 &  23.1 &   39.9 & $32.4\pm2.6$ &  $0.11_{-0.04}^{+0.01}$         & 0 \\
		A2110    &  0.0971 &  23.3 &   30.1 & $55.5\pm7.8$ &  $0.47_{-0.04}^{+0.02}$  & 0 \\
		A2142    &  0.0904 &$<$22.7&   42.8 & $75.1\pm4.7$ &  $2.67\pm0.03$  & 0 \\
		A2199    &  0.0310 &  24.9 &   25.5 & $51.2\pm2.3$ &  $0.63\pm0.01$           & 1 \\
		A2244    &  0.0993 &  22.9 &   37.2 & $38.8\pm2.3$ &  $0.38\pm{0.02}$         & 1 \\
		A2312    &  0.0944 &  24.2 &   44.4 & $21.9\pm2.9$ &  $0.026\pm{0.004}$       & 0 \\
		A2457    &  0.0567 &  22.2 &   36.5 & $21.9\pm6.3$ &  $0.025_{-0.008}^{+0.002}$ & 0 \\
		A2495    &  0.0800 &  23.4 &   29.2 & $46.5\pm6.2$ &  $0.39\pm{0.02}$         & 0 \\
		A2556    &  0.0883 &$<$22.7&   28.4 & $64.4\pm5.1$ &  $0.86\pm0.03$           & 1 \\
		A2566    &  0.0824 &   23.6&   21.5 & $76.4\pm3.0$ &  $0.54\pm{0.06}$         & 0 \\ 
		A2589    &  0.0411 &$<$22.0&   28.8 & $20.8\pm6.4$ &  $0.025\pm0.001$         & 1 \\
		A2597    &  0.0830 &  25.5 &   20.1 & $94.6\pm9.2$ &  $3.01\pm0.03$           & 1 \\
		A2626    &  0.0546 &  23.6 &   26.2 & $39.4\pm2.4$ &  $0.19\pm{0.01}$         & 1 \\
		A2665    &  0.0564 &  23.6 &   25.0 & $31.0\pm5.9$ &  $0.12_{-0.02}^{+0.01}$  & 0 \\
		A3112    &  0.0761 &  25.2 &   21.5 & $56.8\pm3.2$ &  $1.74\pm0.03$           & 1 \\
		A3526    &  0.0099 &  24.0 &   19.1 & $40.5\pm0.5$ &  $0.081\pm0.007$         & 1 \\
		A3528S   &  0.0574 &  24.8 &   27.1 & $27.7\pm5.3$ &  $0.12\pm0.01$           & 1 \\
		A3528N   &  0.0541 &  24.4 &   23.0 & $30.3\pm9.1$ &  $0.11_{-0.03}^{+0.01}$  & 0 \\
		A3571    &  0.0386 &  22.1 &   29.6 & $22.3\pm3.8$ &  $0.072\pm{0.005}$       & 1 \\
		A3809    &  0.0626 &$<$22.6&   27.7 & $45.8\pm3.6$ &  $0.13\pm{0.04}$         & 0 \\
		A4038    &  0.0288 &  22.7 &   24.3 & $50.9\pm1.2$ &  $0.17_{-0.02}^{+0.01}$  & 0 \\
		A4059    &  0.0490 &  24.9 &   27.3 & $33.3\pm1.4$ &  $0.090\pm{0.015}$       & 1 \\
		AWM7     &  0.0173 &$<$21.2&   20.8 & $25.8\pm1.9$ &  $0.046\pm0.002$         & 1 \\
		BVH2007-81& 0.0635 &  23.0 &   31.2 & $18.4\pm8.4$ &  $0.027\pm{0.007}$       & 0 \\
		CYGNUSA  &  0.0561 &  28.1 &   24.3 & $58.8\pm1.5$ &  $1.84\pm0.02$           & 1 \\
		MKW03S   &  0.0453 &  23.7 &   29.6 & $47.6\pm4.4$ &  $0.42\pm0.01$           & 1 \\
	       Ophiuchus &  0.0285 &  22.8 &   25.7 & $36.0\pm2.9$ &  $0.57\pm0.02$           & 1 \\
	       RXCJ0649.3+1801& 0.0639 &$<$22.4&23.7 & $55.2\pm9.6$ &  $0.18\pm0.04$          & 0 \\
	       RXCJ1040.7-7047& 0.0596 & 23.5  &28.6 & $19.6\pm7.2$ &  $0.018\pm{0.002}$      & 0 \\
	       RXCJ1324.7-5736& 0.0186 & 24.2  &21.0 & $39.5\pm2.2$ &  $0.11\pm{0.01}$        & 0 \\
	       RXCJ1539.5-8335& 0.0757 & 23.8  &29.1 & $69.5\pm3.8$ &  $1.73\pm{0.05}$        & 1 \\
	       RXCJ1558.3-1410& 0.0974 & 25.3  &31.0 & $86.1\pm1.3$ &  $1.52\pm{0.23}$        & 0 \\
	       RXCJ1857.6+3800& 0.0538 & 24.3  &27.9 & $15.3\pm2.5$ &  $0.015_{-0.006}^{+0.002}$ & 0 \\
	       RXJ1844.0+4532 & 0.0920 & 26.1  &34.6 & $19.4\pm1.2$ &  $0.048\pm0.012$        & 1 \\
	       RXCJ2218.0-6511& 0.0940 &$<$23.1&20.5 & $32.4\pm5.5$ &  $0.15_{-0.04}^{+0.01}$ & 0 \\
	ZWCL0040.8+2404  &  0.0830 &  23.9 &   17.6 & $73.6\pm2.2$ &  $1.08\pm0.13$  & 0 \\
	ZWCL1742.1+3306  &  0.0757 &  24.1 &   29.4 & $50.7\pm7.5$ &  $0.78_{-0.08}^{+0.04}$  & 1 \\
\hline
\end{tabular}
\small
\begin{tablenotes}
\item
 $a$: redshift of the central BCG;
 $b$: radio luminosity of the central BCG at 1.4 GHz;
 $c$: the offset between the cluster centre and the position of the central BCG;
 $d$: cooling radius with cooling time corresponding to 4 Gyr;
 $e$: X-ray bolometric luminosity within the cooling radius.
	$f$: Note ``1'' represents that $r_{\rm 4Gyr}$ is calculated based on the cooling time profiles in \citet{Cavagnolo09} and ``0'' represents that $r_{\rm 4Gyr}$ is calculated from our analysis.
\end{tablenotes}
\label{tab2}
\end{table*}

\begin{table*}
	\protect\caption{Same as Table \ref{tab2} but for clusters with small offsets ($\le$50 kpc) and small cool cores} 
	\scriptsize
\begin{tabular}{|l|c|c|c|c|c|c|c|c|c|}
\hline 
\hline
		Cluster & $z$ & $P_{\rm 1.4GHz}$ & offset & $r_{\rm 4Gyr}$ & $L_{\rm X,bol}$ & Note \\
		&          & log($P_{\rm 1.4GHz}$/W Hz$^{-1}$)  & (kpc) & (kpc)  & ($10^{41}$ erg/s)    \tabularnewline
\hline 
		3C~129.1 &  0.0222 &  24.3 &   40.9 &  $8.3\pm1.1$ & $2.92_{-0.95}^{+1.00}$  & 0 \\
		A193     &  0.0474 &  23.2 &   35.1 &  $7.6\pm2.3$ & $1.70_{-0.32}^{+0.33}$  & 0 \\
		A376$^{a}$     &  0.0485 &$<$22.1&   28.5 & $15.9\pm2.9$ & $5.18_{-2.42}^{+4.22}$ & 0 \\
		A401     &  0.0714 &$<$22.5&   41.5 &$<$3.0&  $<0.10$                & 0 \\
		A539     &  0.0291 &$<$21.6&   21.7 & $13.0\pm2.1$ &  $6.98_{-0.94}^{+0.50}$ & 1 \\
		A548E    &  0.0396 &$<$22.0&   33.5 & $14.9\pm2.2$ &  $8.23_{-1.84}^{+1.19}$ & 0 \\
		A576     &  0.0411 &   21.8&   26.9 &$<$3.0&  $<0.93$                & 0 \\
		A644     &  0.0705 &$<$22.5&   36.0 &$<$3.6&  $<0.33$                & 0 \\
		A1060    &  0.0128 &  22.4 &   19.9 &  $9.8\pm3.7$ & $0.33\pm0.09$           & 1 \\
		A1314    &  0.0336 &  22.9 &   32.4 &  $6.1\pm1.3$  & $1.60\pm0.14$          & 0 \\ 
		A2107$^{b}$    &  0.0418 & $<$22.0 & 27.3 & $13.8\pm0.5$ & $21.5\pm0.9$      & 1 \\
		A2124    &  0.0661 &$<$22.4&   37.0 & $<3.0$ & $<0.12$  & 0 \\
		A2147    &  0.0354 &  22.6 &   44.3 &$<$3.0& $<0.05$                 & 0 \\
		A2319    &  0.0546 & $<$22.2 &   37.5 &$<$3.0& $<0.06$               & 0 \\
		A2572    &  0.0392 &  23.0 &   30.0 &  $6.8\pm3.3$ & $2.55_{-1.33}^{+1.84}$  & 0 \\
		A2634    &  0.0305 &  24.4 &   44.1 &  $7.2\pm1.9$ & $3.59_{-0.24}^{+0.25}$  & 0 \\
		A2657    &  0.0402 &$<$22.0&   30.5 & $14.1\pm3.5$ & $5.58_{-2.53}^{+1.21}$  & 1 \\
		A2670$^{b}$    &  0.0777 &  22.9 &   37.0 & $13.3\pm4.6$ &  $10.9_{-2.1}^{+1.4}$         & 0 \\
		A2734    &  0.0619 &  23.0 &   34.2 &  $10.4\pm2.9$ & $7.11_{-2.75}^{+4.51}$  & 0 \\
		A2877    &  0.0243 &  21.2 &   24.0 &  $7.7\pm1.1$  & $8.11\pm0.31$          & 0 \\
		A3266    &  0.0602 &$<$22.5&   33.8 &  $5.5\pm1.4$ & $1.01_{-0.47}^{+0.55}$  & 1 \\
		A3330    &  0.0918 &  25.2 &   42.3 &  $7.7\pm2.8$ & $5.28_{-2.27}^{+1.81}$  & 0 \\
		A3341    &  0.0371 &$<$21.9&   31.4 & $12.1\pm6.44$ & $5.22_{-1.68}^{+1.18}$  & 0 \\
		A3391    &  0.0551 &  25.5 &   32.6 &  $3.3\pm0.7$ & $2.35_{-0.36}^{+0.39}$  & 0 \\
		A3395S   &  0.0520 &  25.3 &   31.9 &  $6.8\pm1.4$ & $2.78_{-0.96}^{+1.31}$  & 0 \\
		A3532    &  0.0542 &  24.9 &   31.3 &  $6.8\pm4.6$  & $1.68_{-0.93}^{+2.42}$ & 0 \\
		A3558    &  0.0469 &  22.3 &   32.3 &  $9.5\pm3.5$ & $5.18_{-1.26}^{+1.33}$  & 1 \\
		A3560    &  0.0490 &  24.9 &   41.7 &  $7.0\pm1.9$ & $0.87_{-0.18}^{+0.19}$  & 0 \\
		A3827    &  0.0998 &$<$23.0&   42.5 &  $9.9\pm3.9$ & $0.77_{-0.64}^{+0.66}$  & 0 \\
		A3921    &  0.0938 &$<$22.9&   42.8 &  $11.3\pm2.6$ & $4.57_{-1.79}^{+1.76}$  & 0 \\
		MKW08    &  0.0274 &  21.8 &   31.3 &  $7.2\pm1.6$ & $2.39_{-0.29}^{+0.31}$  & 0 \\
		MS2216.0-0401&0.0942&$<$22.7&  37.9 &  $9.5\pm1.6$ & $5.97_{-1.38}^{+1.43}$  & 0 \\
		RXCJ1614.1-6307&0.0616&$<$22.6&30.4 & $11.3\pm2.7$ & $7.94_{-1.54}^{+1.83}$  & 0 \\
	      SC1327-312 &  0.0501 &$<$22.1&   31.5 &  $8.8\pm2.9$ & $1.61_{-0.77}^{+1.19}$  & 1 \\
	      SC1329-313 &  0.0439 &  21.9 &   32.1 &  $5.9\pm0.9$  & $0.64_{-0.15}^{+0.18}$ & 0 \\
	TrA Cluster      &  0.0494 &$<$22.4&   43.1 & $<$3.0  & $<0.45$              & 0 \\
		ZWCL1215 &  0.0750 &  22.5 &   42.7 &  $6.7\pm1.4$ & $1.95_{-0.73}^{+0.80}$  & 0 \\
\hline
\end{tabular}
\begin{tablenotes}
\item
	$a$: $r_{\rm 4Gyr}$ is slightly larger than 15 kpc, while $L_{\rm X,bol}$ is less than $10^{42}$ ergs/s.
        $b$: $r_{\rm 4Gyr}$ is less than 15 kpc, while $L_{\rm X,bol}$ is larger than $10^{42}$ ergs/s.
\end{tablenotes}
\label{tab3}
\end{table*}

\begin{table*}
	\protect\caption{Same as Table \ref{tab2} but for clusters with large offsets ($>$50 kpc)}
	\begin{tabular}{|l|c|c|c|c|c|c|c|c|c|}
\hline 
\hline
		Cluster & $z$ & $P_{\rm 1.4GHz}$ & offset & $r_{\rm 4Gyr}$ & $L_{\rm X,bol}$ & Note \\
		&          & log($P_{\rm 1.4GHz}$/W Hz$^{-1}$)  & (kpc) & (kpc)  & ($10^{41}$ erg/s)    \tabularnewline
\hline 
		A13      & 0.0905  & 23.7  & 146 & $<$3.0  & $<$0.20              & 0 \\
		A119     & 0.0445  &$<$22.0& 63.5 & $2.7\pm0.7$  & $0.16_{-0.08}^{+0.11}$ & 0 \\
		A754     & 0.0543  &$<$22.2& 65.0&$<$3.0& $<0.10$                 & 0 \\
		A1142    & 0.0368  & 23.1  & 114 & $9.4\pm2.1$  & $1.26\pm0.17$           & 0 \\
		A1185    & 0.0347  &$<$21.8& 129 & $5.9\pm2.3$  & $0.28_{-0.09}^{+0.11}$  & 0 \\
		A1367    & 0.0211  & 24.8  & 147 & $5.1\pm1.7$  & $0.80\pm0.09$           & 0 \\
		A1656    & 0.0239  & 23.4  &  54  & $4.1\pm1.5$  & $0.78_{-0.12}^{+0.15}$ & 0 \\
		A2061    & 0.0791  &$<$22.6& 254 & $7.7\pm1.0$  & $4.29_{-0.79}^{+0.78}$  & 0 \\
		A2256    & 0.0594  &$<$22.3& 117 & $<$3.9  & $<0.50$              & 0 \\
		A2384S   & 0.0956  & 24.1  & 541 & $24.4\pm1.6$ & $22.8_{-3.4}^{+3.6}$      & 0 \\
		A2440    & 0.0902  &$<$22.7& 443 & $29.2\pm3.8$ & $27.7_{-9.2}^{+11.9}$     & 0 \\
		A3376    & 0.0456  & 24.3  &  99  & $5.1\pm1.6$  & $0.59_{-0.14}^{+0.17}$ & 0 \\
		A3627    & 0.0182  & 25.4  & 110 & $5.4\pm0.3$  & $4.42\pm0.71$  & 0 \\
		A3667    & 0.0557  &$<$22.5& 114 & $<$3.6  & $<0.27$              & 0 \\
		A3744    & 0.0384  & 25.3  &  95  & $6.1\pm1.2$ & $0.75_{-0.25}^{+0.31}$  & 0 \\
		A4067    & 0.100   &$<$23.0& 235 & $28.2\pm2.8$ & $47.4_{-14.9}^{+17.6}$      & 0 \\
		AS463    & 0.0399  & 25.3  &  75  & $7.9\pm1.2$  & $1.66_{-0.30}^{+0.31}$ & 0 \\
\hline
\end{tabular}
\label{tab4}
\end{table*}

\section{Results}
\label{Results}

\subsection{Cool Core Size vs. Radio Luminosity}
\label{size}
\begin{figure}
\begin{center}
\hbox{\hspace{-10px}
\includegraphics[scale=0.44]{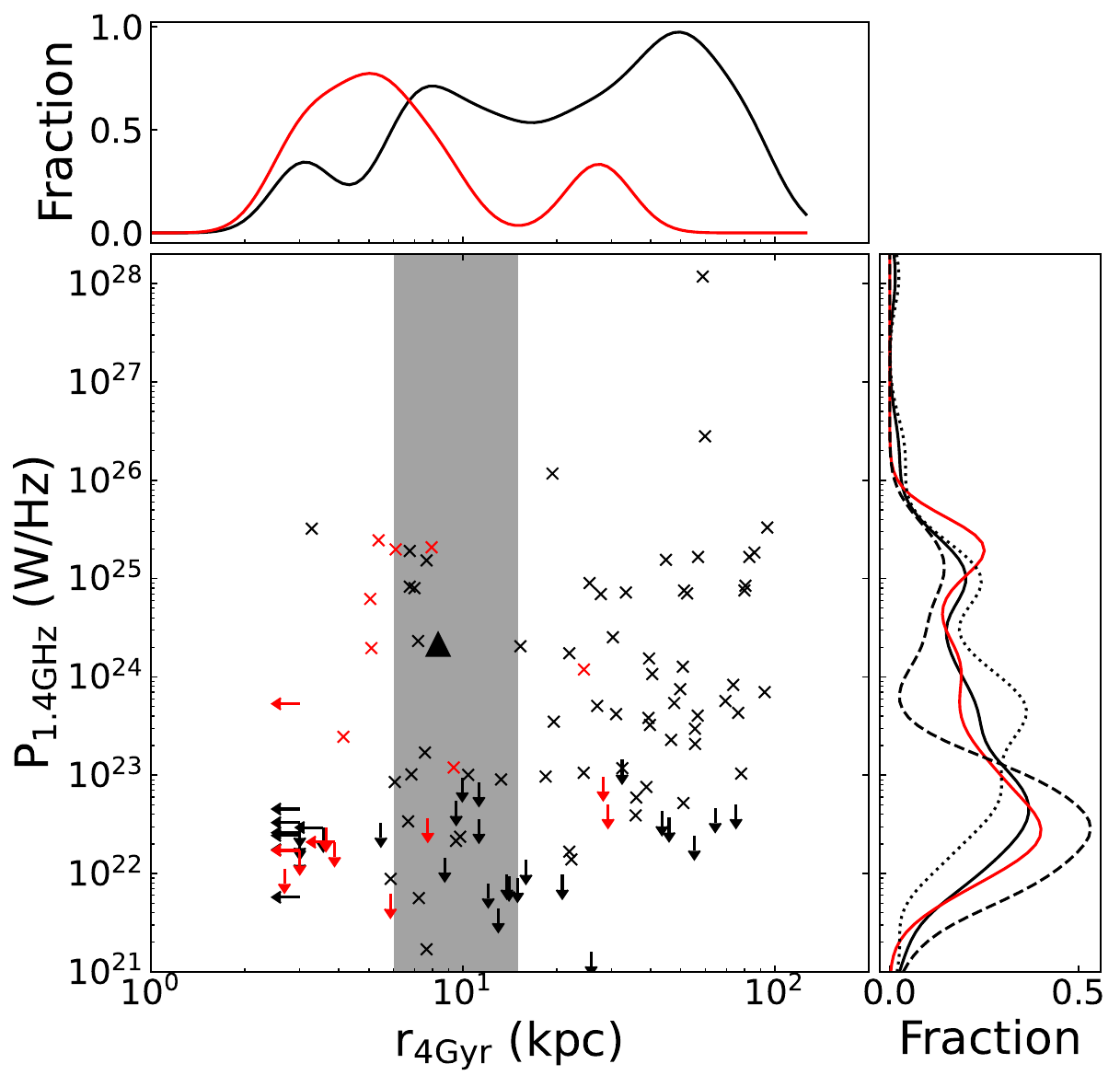}
}
\caption{
	Cool core size vs. radio power of the central BCG at 1.4 GHz. The shaded region shows the systems with 
	cool core sizes between 6 to 15 kpc. On the left are systems with coronae and on the right are systems with large cool 
	cores. Systems with small offsets are in black and those with large offsets are in red. 
	The triangle symbol represents the cluster 3C~129.1 (see section \ref{Case}). The Gaussian KDEs for cool core size and radio power 
	distributions of clusters with small offsets (black solid) and large offsets (red solid) are shown. 
	The Gaussian KDEs for the radio power distribution for clusters with small offsets and with small cool cores 
	($r_{\rm 4 Gyr} \le$ 15 kpc) and large cool cores ($r_{\rm 4 Gyr} >$ 15 kpc) are shown by the dashed line and the dotted line 
	respectively. For clarity purpose, we did not plot the error bar for the cool core size.
}
\label{rcPr}
\end{center}
\end{figure}

We plotted the cool core size vs. radio luminosity of the central BCG at 1.4 GHz
in Fig.~\ref{rcPr}. The clusters with transition CCs are shown 
in the shaded region, while corona clusters and LCC clusters are on the left and right, respectively. The top panel of Fig.~\ref{rcPr} 
shows the CC size distributions for galaxy clusters with small and large offsets respectively from the Gaussian kernel density estimates (KDEs).
We looked at the systems with small offsets ($\le$50 kpc) between the central BCG and the X-ray cluster centre,
and about 40\% of them (36 out of 91) have small CCs. While for the systems with large offsets ($>$50 kpc), 
We found most of them have small CCs, i.e., only 3 out of 17 clusters have large CCs,
and the largest cooling radius is $\sim29$ kpc (A2440). This comparison of two subsamples with different offsets suggests 
that cluster mergers or sloshing can disrupt LCCs efficiently to reduce the cool core size. Mergers trigger bulk motion or 
sloshing of cool cores that spreads the metal-rich gas in a larger volume.

We checked the relation between the CC size and radio luminosity of the BCG for clusters with small offsets.
There is a weak correlation between the CC size and radio power for clusters
with large CCs, with a correlation coefficient of $0.29\pm0.12$ obtained using the tool 
linmix\footnote{https://github.com/jmeyers314/linmix}. While there is no correlation 
between them for clusters with small CCs, with a correlation coefficient consistent with zero.

We also showed the radio power distributions for the relaxed and disturbed systems in the right panel of Fig.~\ref{rcPr}.
We found that the fraction of radio weak systems ($\sim47\%$ with $P_{1.4 \rm GHz}<10^{23}$ W Hz$^{-1}$)
in disturbed clusters is comparable to that (45\%) in relaxed clusters, though the total number of disturbed
clusters is much smaller in our sample.
For clusters with small offsets, we plotted the radio power KDEs for small CCs and large CCs, respectively.
The large CC systems show a relatively continuous distribution of radio power.
However, most of small CC systems (71\%) have a weak radio AGN with radio power $P_{1.4 \rm GHz}<10^{23}$ W Hz$^{-1}$ and
there is a lack of small CC systems with intermediate radio power from 
$\sim2\times10^{23}$ W Hz$^{-1}$ to $\sim 2\times10^{24}$ W Hz$^{-1}$. By contrast, 19 large CC clusters fall within this
radio power range. 
For systems with strong radio AGNs, there is a total of 10 small CC clusters,  
while there are 40 large CC clusters. 
This may suggest different duty cycles (defined as the fraction of the time when the galaxy hosts a radio-loud AGN, 
see more details on section \ref{duty}) of radio AGN feedback processes in large CC and small CC clusters.
In large CCs, the radio AGN stage is more continuous.
In small CCs, the radio AGNs spent a much shorter time in the radio strong stage than in the radio weak stage.

\subsection{Cool Core Luminosity vs. Radio Luminosity}
\begin{figure}
\begin{center}
\hbox{\hspace{-8px}
\includegraphics[scale=0.43]{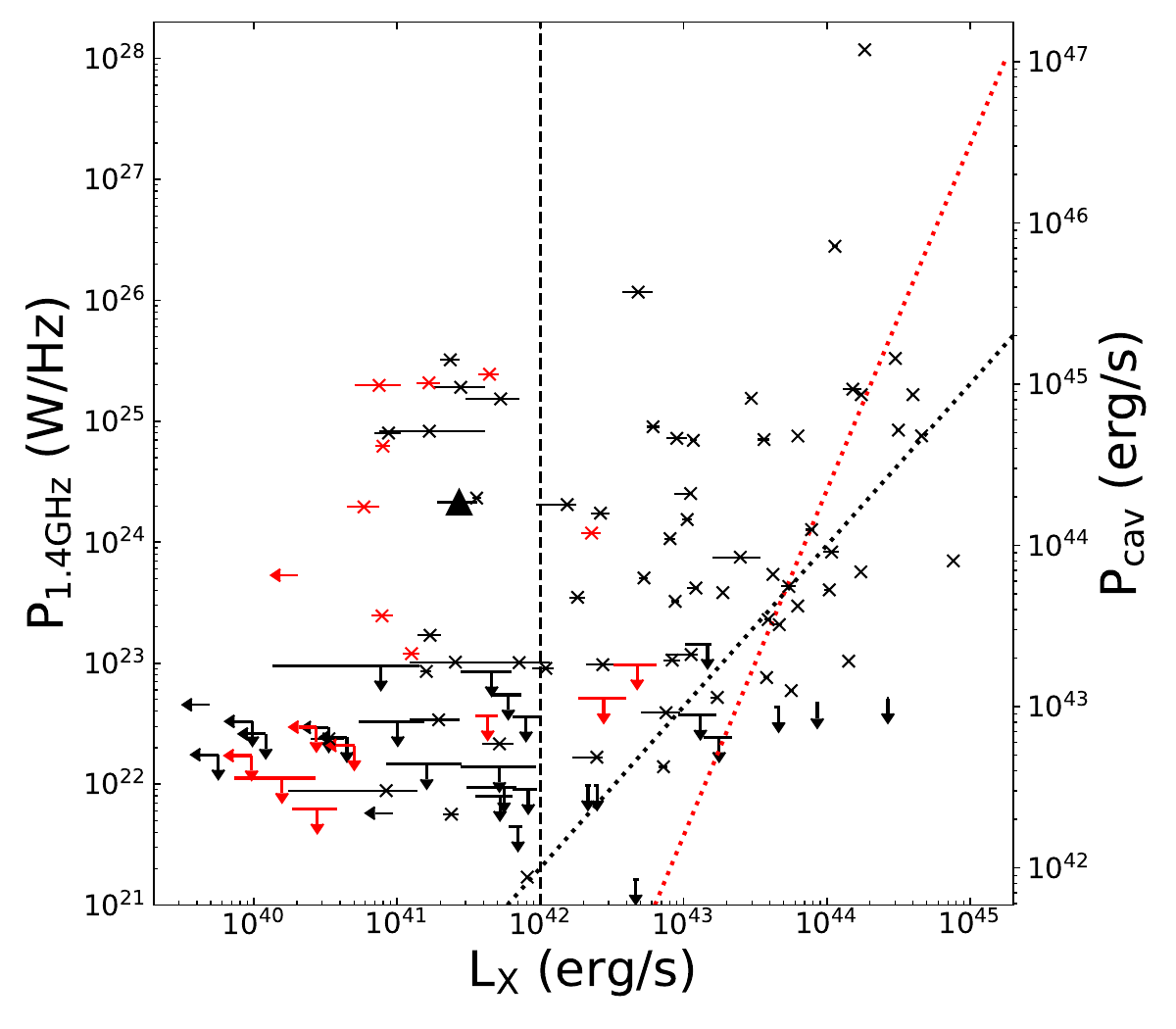}
}
\caption{
	Bolometric X-ray luminosity within the cooling radius vs. radio power of the central BCG at 1.4 GHz. The dashed line 
	($L_{\rm X,bol}=10^{42}$ erg s$^{-1}$) separates the large cool cores and small cool cores. Data points in red are for the 
	systems with large offsets (greater than 50 kpc). The triangle symbol represents the cluster 3C~129.1 (see section \ref{Case}).
	The y-axis on the right is the cavity power estimated using the relation from \citet{Cavagnolo2010}.
	The dotted lines show equality of bolometric X-ray luminosity and cavity power, which is estimated 
	using the relations from \citet{Birzan08} (red dotted line) and \citet{Cavagnolo2010} (black dotted line), respectively. 
}
\label{LxPr}
\end{center}
\end{figure}

We then examined BCGs in the plane of the bolometric X-ray luminosity within the cooling radius $r_{4\rm Gyr}$ vs. radio luminosity at 1.4 GHz
in Fig. \ref{LxPr}. 
As shown in Fig.~\ref{LxPr} we divided all CCs into two classes, small CCs and large CCs
with a dividing bolometric X-ray luminosity of $10^{42}$ erg~s$^{-1}$. This is relatively
consistent with the previous study, where they used a dividing 
CC luminosity of $4\times10^{41}$ erg s$^{-1}$ in 0.5-2.0 keV band \citep{Sun09},
considering that the conversion factor from 0.5-2 keV luminosity to bolometric luminosity for low-temperature corona gas is $\sim$ 2.
In this way, the division of large CCs and small CCs based on the cool core luminosity is almost consistent with the 
division based on the CC size. We found three clusters in different classes with these two different criteria, A2107, A2670, and A376, with the former
two having CC luminosities of $>10^{42}$ erg s$^{-1}$ while having CC sizes of $<15$ kpc, and
the last one having a CC luminosity of $\sim5\times10^{41}$ erg s$^{-1}$ while having 
a CC size of $\sim16$ kpc. On the other hand, their exact classification has little impact on conclusions of our work.

In Fig. \ref{LxPr}, the large CC clusters show a weak correlation between the CC luminosity and 
the radio luminosity of the BCG. More radio luminous sources generally reside in the larger cool cores,
although with a large scatter. Using the tool linmix,
we calculated a positive correlation coefficient of $0.38\pm0.12$ between the radio luminosity and CC luminosity of the large CC clusters.
We also plotted the kinetic power based on the $P_{radio}- P_{cavity}$ relation from \citet{Birzan08} and \citet{Cavagnolo09}.
For most of the large CC clusters, the inferred kinetic power exceeds the cooling luminosity. 
Notice that the cooling luminosity here is estimated within the cooling radius with the cooling time corresponding to 4 Gyr.
Using a large cooling radius, e.g., with a typical cooling time of 7.7 Gyr, can have a higher cooling luminosity.
In comparison, the small CC clusters do not show a correlation between the radio power and the CC luminosity,
with a correlation coefficient consistent with zero.
The radio luminous corona systems show that the inferred kinetic power could greatly exceed the CC luminosity,
suggesting that most of the energy released from central AGN needs to be deposited outside of the corona in order for it to survive.

Radio and X-ray observations have shown that radio jets created by the central AGN can drill through the hot atmosphere of the central 
BCG and deposit a large amount of energy into the ambient gas at large radii \citep[e.g.,][]{Smith2002,Lane2004,Sun2005,OSullivan2010}. 
The large CCs could be disrupted and transform into the coronae \citep[e.g.,][]{Soker2016}.
However, we did not observe many clusters with strong radio AGNs ($P_{1.4 \rm GHz} > 2\times10^{24}$ W~Hz$^{-1}$)
while having CC luminosities between $10^{42}$ to $10^{43}$ erg s$^{-1}$, or a CC size between about 10 to 20 kpc
in Fig. \ref{rcPr} and \ref{LxPr}.
In fact, there are a total of 17 clusters with $P_{1.4 \rm GHz} > 2\times10^{24}$ W~Hz$^{-1}$ and with large CCs.
The fraction of these clusters decreases with the CC luminosity, e.g., from $\sim47\%$ for clusters 
with $10^{44}<L_{\rm X,bol}<10^{45}$ erg~s$^{-1}$ to $\sim24\%$ for clusters with $10^{42}<L_{\rm X,bol}<10^{43}$ erg~s$^{-1}$.
Therefore, the low fraction of such clusters, with $10^{42}<L_{\rm X,bol}<10^{43}$ erg~s$^{-1}$ and strong radio AGN, 
suggests that if the large CCs can be disrupted by strong 
radio AGN and turned into the coronae, the time spent during this stage should be short.
However, we need to note that the number of clusters in this range is quite small, and there is a possibility that our sample does not have many such clusters by chance.
A statistically large, complete cluster sample will help to check whether our interpretation is true or not.

\subsection{Circumnuclear X-ray Cool Cores}
\begin{figure}
\begin{center}
\hbox{\hspace{-0.3cm}
\includegraphics[scale=0.49]{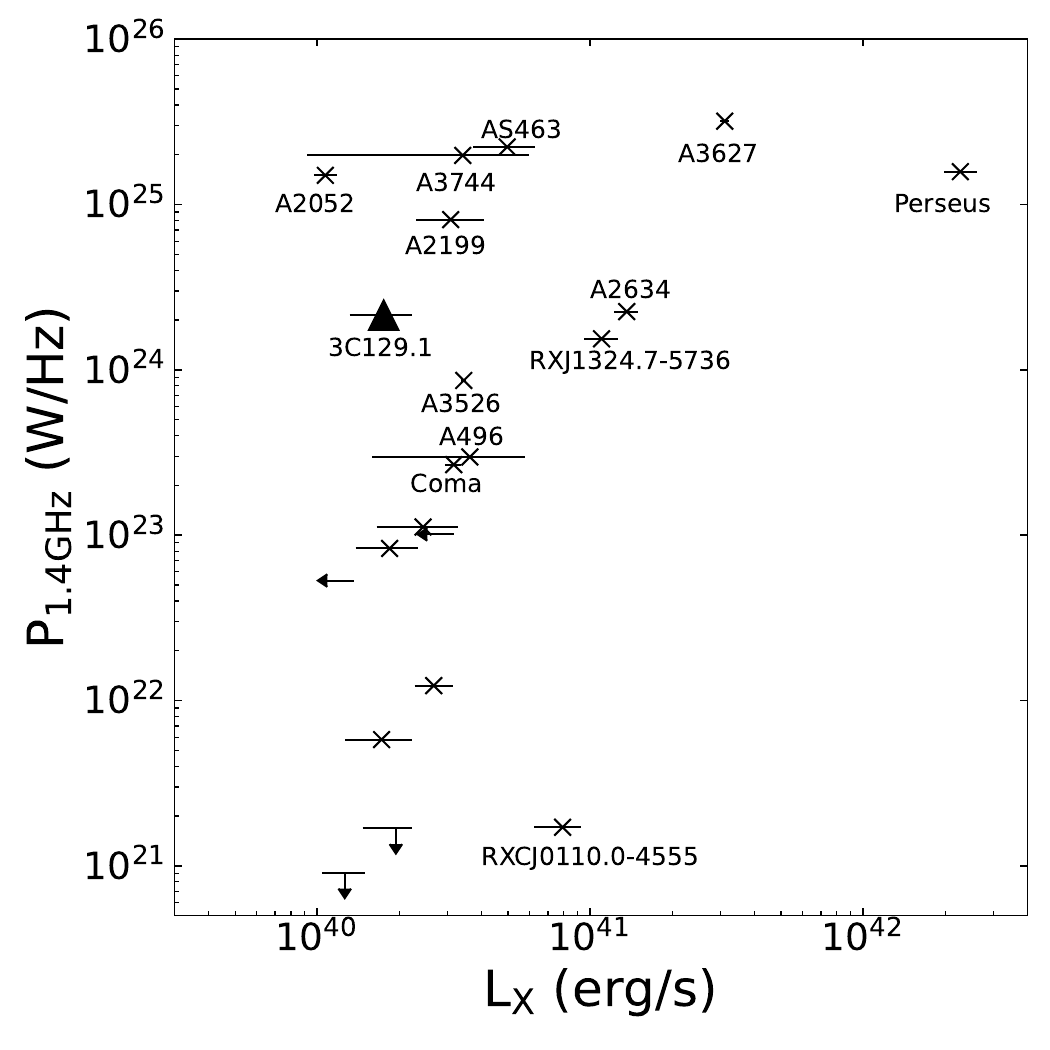}
}
\vspace{-0.4cm}
\caption{
	Bolometric X-ray luminosity within 1 kpc vs. radio power of the central BCG at 1.4 GHz for clusters within redshift 
	$z<0.04$ in our sample. The triangle symbol represents the cluster 3C~129.1.
	There is a lack of clusters with both strong circumnuclear X-ray cool cores ($L_{\rm X} > 10^{41}$ erg/s) and 
	weak radio AGN ($P_{\rm 1.4 GHz} < 10^{24}$ W Hz$^{-1}$). 
}
\label{Lx1kpc}
\end{center}
\end{figure}

As AGN feeding is a phenomenon happening in the very centre of the BCG, we also explore the relationship between the circumnuclear X-ray luminosity vs. the 
radio luminosity of the BCG (Fig.~\ref{Lx1kpc}). We defined the circumnuclear X-ray luminosity in this study as the bolometric X-ray 
luminosity enclosed within 1 kpc of the central BCG. We extracted the source spectrum from a circular region centred on the
BCG with a radius of 1 kpc, and the background spectrum within an annulus from 1 kpc to 2 kpc.
In order to robustly measure the luminosity within the central 1 kpc, we only studied clusters at $z <$ 0.04, where 1 kpc $> 1.26''$ to be 
resolved by \chandra{}, and with more than 60 counts in the $0.5-7.0$ keV band within 1 kpc after background subtraction for statistical constraints.
There are a total of 21 galaxy clusters which satisfy these criteria. To measure the circumnuclear X-ray luminosity, we fitted the spectrum
with an absorbed thermal component for the hot gas, and a power-law component to account for the possible non-thermal emission from the central AGN.
We found that there are no systems with low radio power ($P_{1.4\rm GHz} < 10^{24}$ W Hz$^{-1}$) and strong 
circumnuclear X-ray luminosity ($L_{\rm X} > 10^{41}$ erg s$^{-1}$). 
In other words, BCGs with strong circumnuclear X-ray CC ($L_{\rm X} > 10^{41}$ erg s$^{-1}$) within 1 kpc are always associated 
with strong radio AGN ($P_{1.4\rm GHz} > 10^{24}$ W Hz$^{-1}$).
This may suggest that high circumnuclear X-ray luminosity in galaxy cluster may be used as a probe of radio AGN. 
However, this result needs to be examined with a large, representative sample.

We also estimated the offset between the BCG and the X-ray peak for these 21 clusters. In general, the offsets are relatively
small, within 0.6 kpc for all but five clusters in our sample. The largest offset is $\sim2$ kpc for the Ophiuchus cluster, 
whose circumnuclear X-ray luminosity is small, with an upper limit of $1.4\times10^{40}$ erg s$^{-1}$.
A high-resolution \chandra\ X-ray image reveals that the peak of cluster CC is displaced from the central BCG, which 
is probably the reason for the lack of strong AGN activity in both radio and X-ray \citep{Werner2016}. Recently, using low-frequency radio data,
\citet{Giacintucci2020} found a giant cavity filled with diffuse radio emission with a steep radio spectrum, which could be
due to an extraordinarily powerful past AGN outburst. The steep radial density and temperature gradient of the Ophiuchus CC may help it
survive disruption by AGN jets piercing the CC and depositing energy right outside the core. 
The other three clusters with more than 0.6 kpc offset are 2A0335+096, A496, and AWM7.

\section{A Detailed Case Study of 3C~129.1}
\label{Case}
In this section, we present a detailed analysis of a corona system with a strong radio AGN. The 3C~129.1 cluster, 
with a temperature of $\sim5.6$ keV \citep{Ikebe02}, contains two strong radio sources, i.e., 3C~129 and 3C~129.1. 3C~129.1 
(or WEIN 051) is the BCG of this cluster. 3C~129.1 was observed by \chandra\ for 9.5 ksec in 2001 (ObsID 2219).
The previous studies showed that the system has a heated core and several candidates of ultraluminous X-ray sources (ULXs) \citep{Krawczynski2002, Krawczynski2003}. 
At the very centre, 3C~129.1 hosts a small, extended X-ray source and it is most likely a thermal corona \citep{Sun2007}. 
In fact, in this work we found the cluster core is heated, with an elevated metal abundance (see section \ref{3c129_results}), which was proposed to be a signature
of a cool core remnant \citep[e.g.,][]{Rossetti2010}.
Here, we used new \chandra\ observations to study this candidate of a CC remnant.
\begin{figure*}
\hbox{\hspace{-15px}
\begin{tabular}{lll}
\includegraphics[scale=0.25]{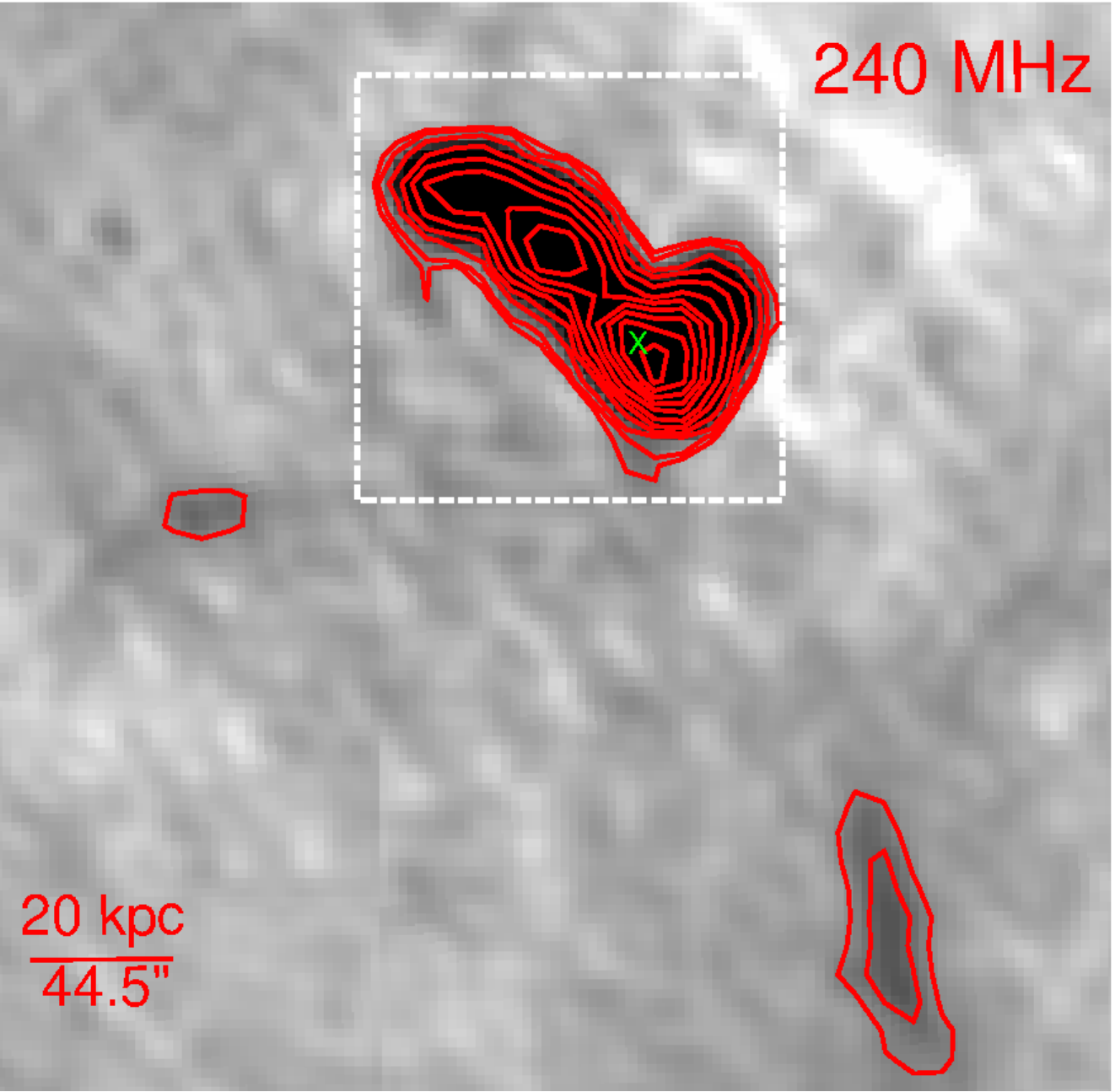}
&
\hspace{-5px}
\includegraphics[scale=0.25]{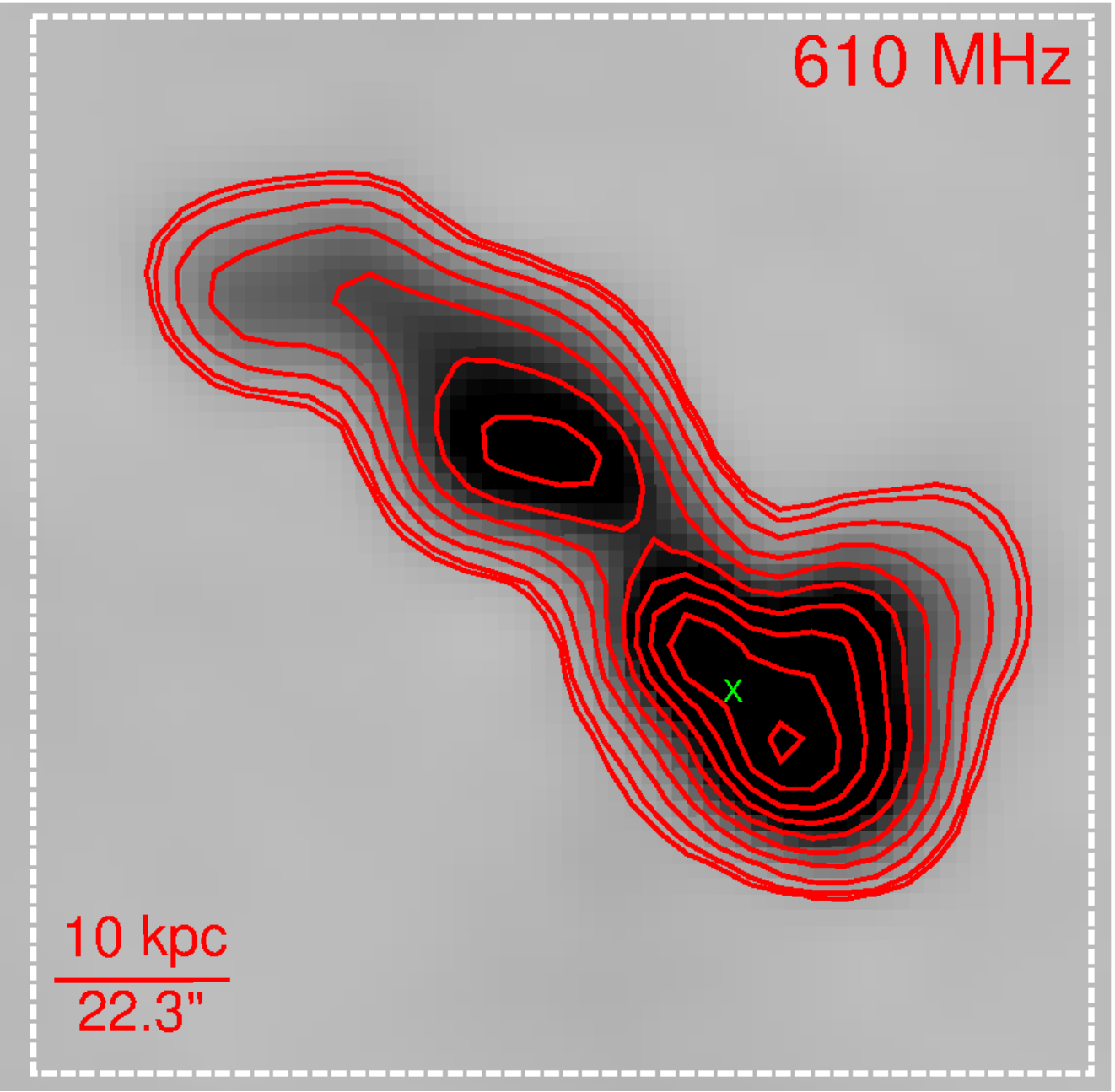}
&
\hspace{-5px}
\includegraphics[scale=0.25]{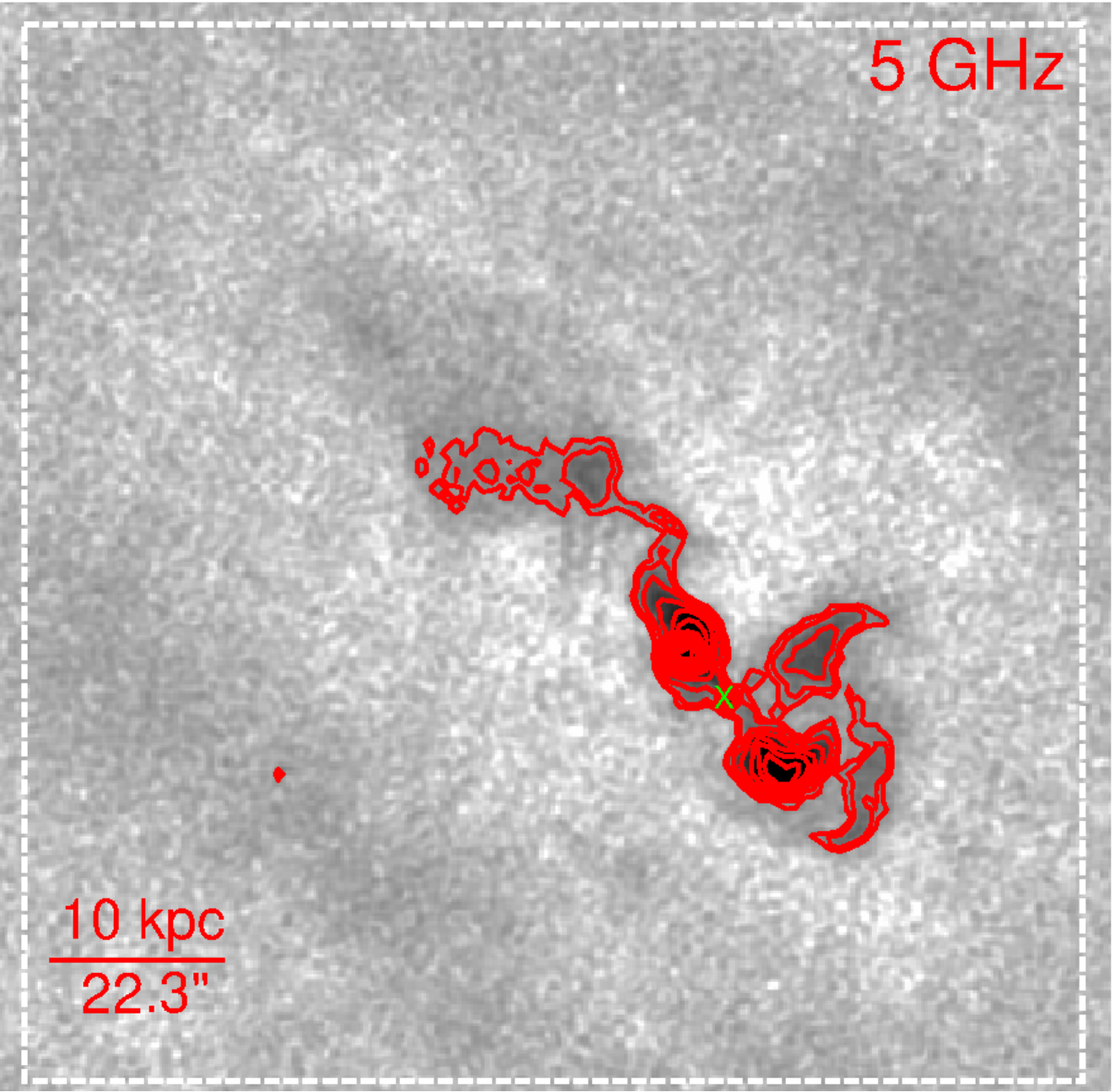}
\end{tabular}
}
\caption{
	\gmrt\ 240 MHz (left), 610 MHz (middle), and \vla\ 4.76 GHz (right) images with an rms level of 1300, 300, and 25.1 $\mu$Jy/beam ($1\sigma$)
	and contours of 3C~129.1. The \gmrt\ 204 MHz and 610 MHz images have a resolution of 
	13.1\arcsec$\times$10.8\arcsec\ and 8.0\arcsec$\times$7.1\arcsec, respectively. The beam size of \vla\ image is 1.25\arcsec.
	The \gmrt\ 240 MHz contours start from 4 $\sigma$ to 0.4 Jy/beam. The \gmrt\ 610 MHz
	contours start from 30 $\sigma$ to 1000 $\sigma$. The \vla\ 4.76 GHz contours start from 4 $\sigma$ to 200 $\sigma$. All contours used
	the square-root spacing. In the left panel faint diffuse emission is detected to the southwest 
	($\sim2.5\arcmin-4.3\arcmin$ from the nucleus), and to the southeast ($\sim2\arcmin-2.8\arcmin$ from the nucleus).
	The green cross indicates the position of the central BCG.
	The white boxes are the same in all panels with a size of 2.25\arcmin$\times$2.25\arcmin.
}
\label{radio}
\end{figure*}

\subsection{\chandra\ Data Analysis}
3C~129.1 was observed by \chandra\ for $\sim80$ ksec split into two observations (Obs. IDs 1957 and 19965, PI: Sun) in December 2016 
with the Advanced CCD Imaging Spectrometer (ACIS). Both observations (Aimpoint on chip S3) were taken in 
Very Faint (VFAINT) mode and centred on the cool core. The details of the \chandra\ observations are 
summarized in Table \ref{tab_obs}. There was one ACIS-I observation taken in 2001 with a short exposure of
$\sim9.5$ ksec. Another ACIS-S observation was taken in 2000, however, it is centred on the galaxy 3C~129. 
In this study, we focus on the newer ACIS-S data.
The data analysis followed the standard process (see section \ref{DataAnalysis}). 
To improve background filtering we set {\tt CHECK\_VF\_PHA} as yes to remove bad events that are likely associated with cosmic rays.
We checked that there were no strong background flares for any observations and the resulting cleaned exposure time is shown in 
Table \ref{tab_obs}. The CALDB blank sky background was used in our data analysis.
We estimated the X-ray sky background residuals using background models including two APEC components with zero redshift and Solar abundance and
one power-law component.

The 3C~129.1 cluster lies close to the Galactic plane, so the Galactic absorption is important to our spectral analysis.
In our spectral analysis, we allowed the absorption column density to be free (see Appendix \ref{appendix}).

\begin{table}
\protect\caption{\chandra\ Observations of 3C~129.1 (PI: Sun)}
\begin{tabular}{|c|c|c|c|}
\hline 
ObsID & Date Obs & Total Exp    & Cleaned Exp \\
      &          & (ks)         & (ks)        \tabularnewline
\hline 
19567 & 2016 Dec 20 & 15.1 & 14.3 \tabularnewline
19965 & 2016 Dec 21 & 64.2 & 63.5 \tabularnewline
\hline 
\end{tabular}
\label{tab_obs}
\end{table}

\subsection{Radio Data Analysis}
The Giant Metrewave Radio Telescope (\gmrt) observations at 240 MHz and 610 MHz were carried out on 2002 May 19 and May 24, respectively.
The data were collected in the standard spectral line mode with a spectral resolution of 125 kHz.
The visibility data were converted to FITS and analyzed using the NRAO Astronomical Image Processing System package (AIPS) \citep{Lal2005}.
The standard flux density calibration source was observed in the beginning as an amplitude calibrator and also to estimate and correct for the bandpass shape.
The phase calibration source was observed once every 35 min.
The error in the estimated flux density, both due to calibration and systematic, is less than $\sim$ 5\%.
The data suffered from intermittent radio frequency interference. Thus, in addition to normal editing of the data, the channels affected due to radio frequency 
interference were identified and edited, after which the central channels were averaged using AIPS task `SPLAT'.
To avoid bandwidth smearing, 6.75 MHz of clean band at 240 MHz was reduced to 6 channels of 1.125 MHz each, and 13.5 MHz of clean band 
at 610 MHz was averaged to give 3 channels of 4.5 MHz each.
While imaging, 49 facets, spread across $\sim$2$^\circ$ $\times$ 2$^\circ$ field were used at 240 MHz, and 9 facets covering slightly less 
than 0.7$^\circ$ $\times$ 0.7$^\circ$ field were used at 610 MHz to map each of the two fields using AIPS task `IMAGR'.  
We used `uniform' weighting and the 3-D option for `W'-term correction throughout our analysis.
After 2-3 rounds of phase self-calibration, a final amplitude and phase self-calibration was made to get the final image.
At each round of self-calibration, the image and the visibilities were compared to check for the improvement in the source model.
The final maps were stitched together using AIPS task `FLATN' and corrected for the primary beam shape of the GMRT antennas using AIPS task `PBCOR'
\citep[see also][]{Lal2019}. The final image at 240 MHz has a resolution of 13.1\arcsec$\times$10.8\arcsec and an rms noise level (1 $\sigma$) of 1.3 mJy/ beam.
The final image at 610 MHz has a resolution of 8.0\arcsec$\times$7.1\arcsec\ and an rms noise level of 0.3 mJy/ beam.

The Very Large Array (\vla) image at 4.76 GHz was downloaded from NRAO \vla\ archive\footnote{https://www.vla.nrao.edu/astro/nvas/}. 
The observation was taken on 2000 Feb 8 with B configuration. The image was generated using the \vla\ pipeline in AIPS. The
beam size is 1.25\arcsec\ and the rms level is 25.1 $\mu$Jy/beam.

The radio images of 3C~129.1, from \gmrt\ at 240 MHz and 610 MHz, and 
\vla\ at 5 GHz, are shown in Fig. \ref{radio}. The radio emission of 3C~129.1 is relatively compact
with a radio power at 1.4 GHz of $\sim2\times10^{24}$ W Hz$^{-1}$.
\subsection{Properties of 3C~129.1 ICM and Cool Core Remnant}
\label{3c129_results}

\begin{figure*}
\hbox{\hspace{-15px}
\begin{tabular}{lll}
\includegraphics[scale=0.22]{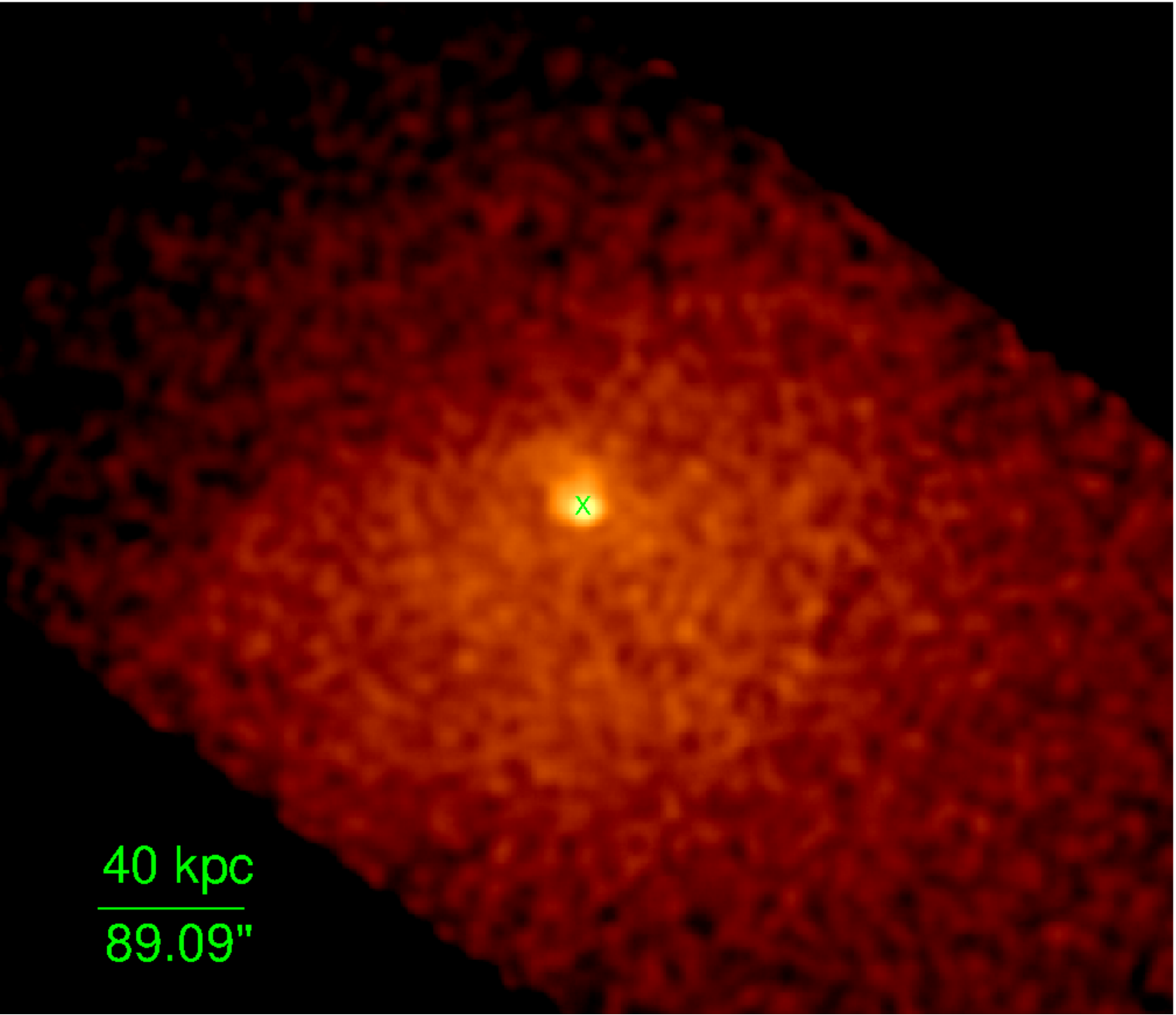}
&
\hspace{-12px}
\includegraphics[scale=0.22]{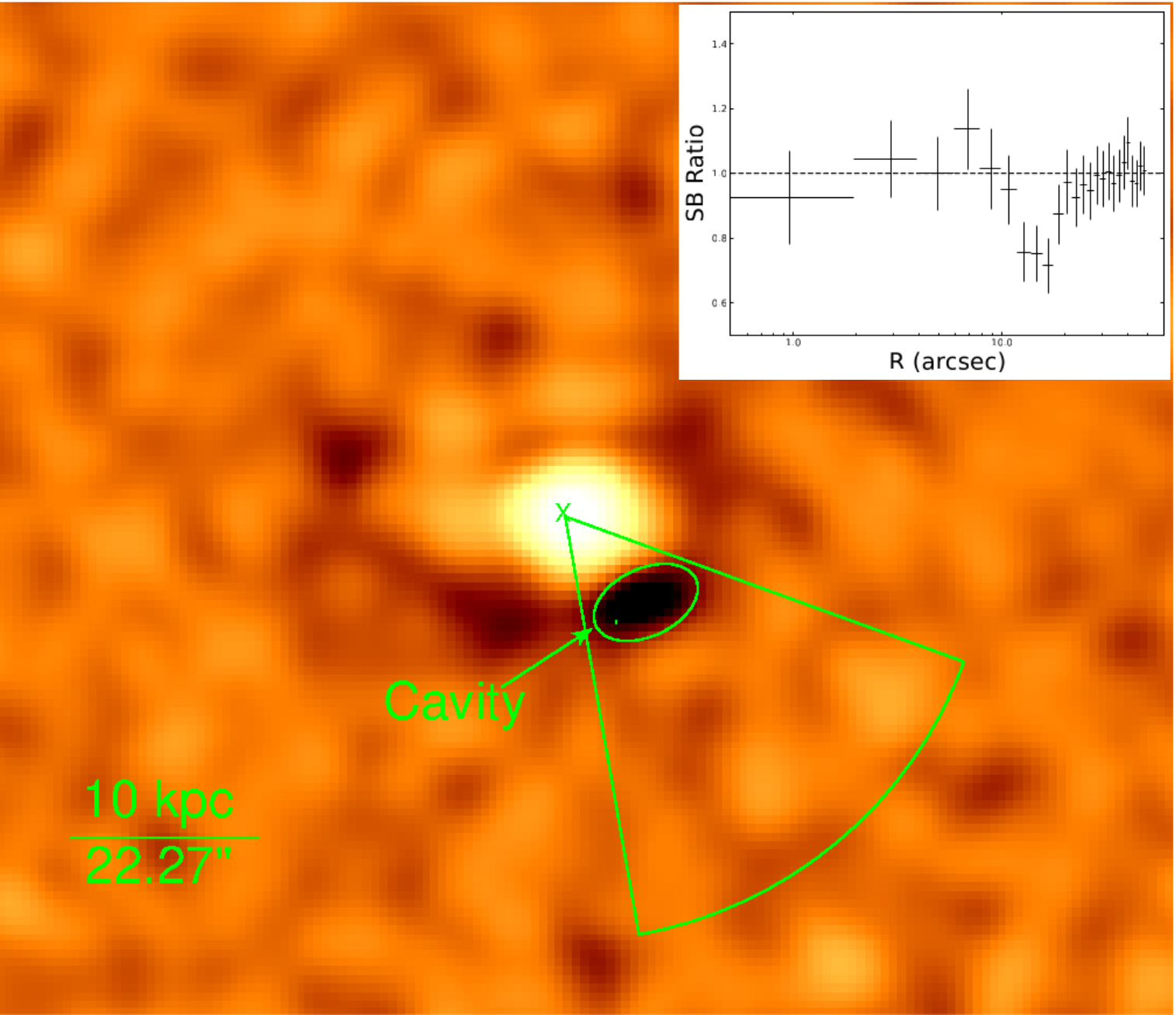}
&
\hspace{-12px}
\includegraphics[scale=0.22]{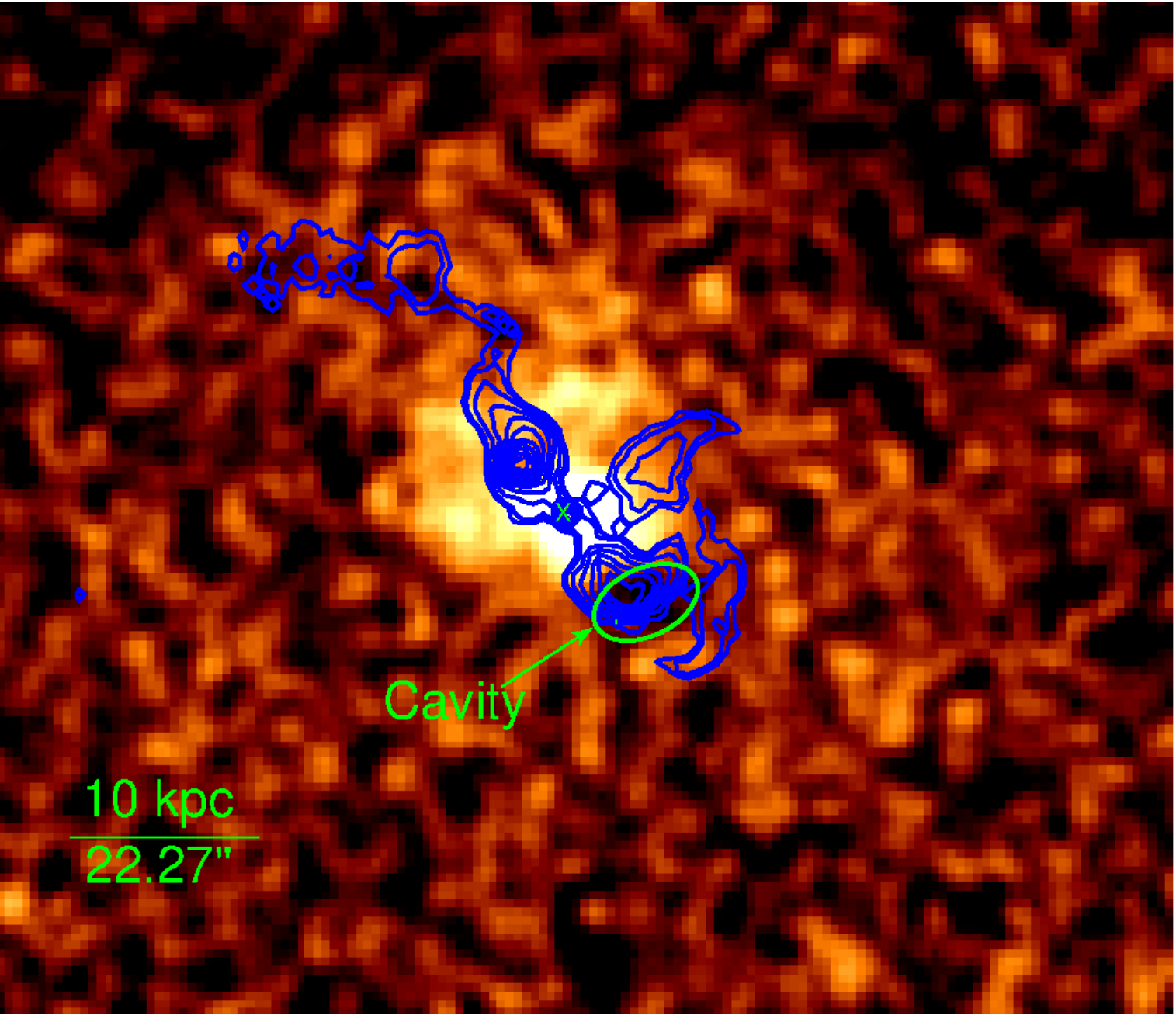}
\end{tabular}
}
\caption{Left: The combined background-subtracted, exposure-corrected \chandra\ images of 3C~129.1 in the 
	0.7-4.0 keV band, smoothed by a 10-pixel Gaussian kernel. The central bright region is a small
	cool core. The X-ray point sources have been removed and filled with the local background.
	Middle: The unsharp-masked image created by subtracting the 0.7-4.0 keV background-subtracted, exposure-corrected
	image smoothed with 10-pixel Gaussian kernel from the image smoothed with 3-pixel Gaussian kernel. 
	The ellipse indicates the X-ray cavity. The wedge is used for surface brightness analysis for the detection
	of the cavity. The inserted image is the ratio of the surface brightness along the wedge to the azimuthally-averaged
	surface brightness. There is a clear surface brightness depression from $\sim10\arcsec$ to $\sim20$\arcsec.
	Right: The zoomed-in \chandra\ image as in the left panel smoothed with a 3-pixel Gaussian kernel,
	overlaid with the \vla\ 5 GHz radio contours (blue). The contour levels are the same as in Fig. \ref{radio}. 
	The X-ray cavity overlaps with the radio lobe.
	The green crosses in all panels indicate the position of the central BCG.
	}
\label{image}
\end{figure*}

Fig. \ref{image} (left) shows the combined background-subtracted, exposure-corrected \chandra\ images of 3C~129.1 in the
0.7-4.0 keV band, smoothed by a 10-pixel Gaussian kernel. The central bright region shows the corona with a 
size of less than 10 kpc. 
We used unsharp masking to determine the size and position of X-ray cavity. Fig. \ref{image} (middle) shows the unsharp-masked
image generated by subtracting the 0.7-4.0 keV background-subtracted, exposure-corrected image smoothed
with 10-pixel Gaussian kernel from the image smoothed with 3-pixel Gaussian kernel.
A potential X-ray cavity is shown with the ellipse.  
We then extracted the azimuthally-averaged surface brightness profile and the surface brightness profile 
in the wedge along the southwestern direction. As shown in the inserted figure in the middle panel of
Fig. \ref{image}, there is a clear surface brightness depression from $\sim10\arcsec-20\arcsec$, corresponding to the position 
of X-ray cavity. 
Based on the surface brightness, we calculated the significance of the depression of $\sim3\sigma$ within the cavity.
Fig. \ref{image} (right) is the zoom-in \chandra\ image overlaid with the \vla\ 5 GHz radio contours.
The bright core is a little bit more extended toward the northeast. The X-ray cavity overlaps with 
the southwestern radio lobe.

\begin{figure*}
\begin{center}
\hbox{\hspace{-25px}
\includegraphics[scale=0.55]{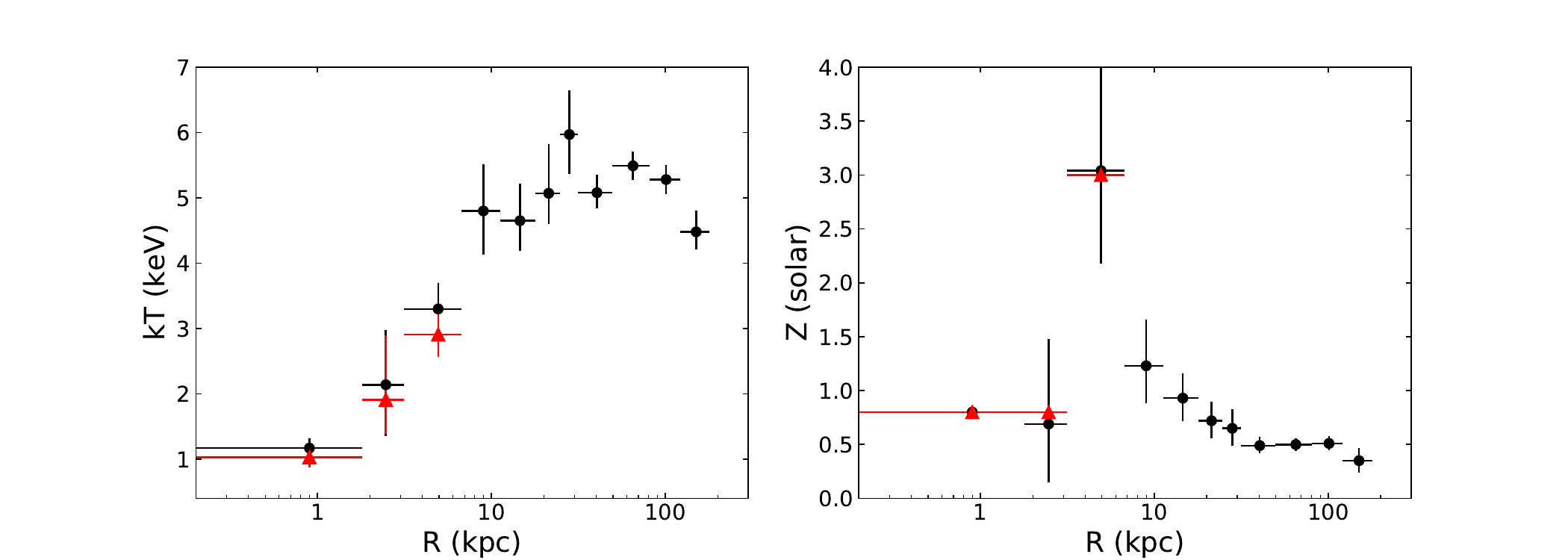}
}
\vspace{-0.2cm}
\caption{
	The azimuthally-averaged radial profiles of the temperature (left) and abundance (right) of the cluster 3C~129.1. 
	The black circles represent the projected temperature and abundance values, and the red triangles represent the deprojected values.
}
\label{Tprofile}
\end{center}
\end{figure*}
To obtain temperature profiles, we extracted the spectra from a set of 11 circular annuli centred on the BCG to a radius of $\sim180$ kpc 
for two observations using the CIAO tool ``specextract''.
The spectra from the same region from two observations were fitted simultaneously to an absorbed thermal APEC model, i.e., TBABS*APEC,
in the energy range $0.5-7.0$ keV, with the fixed background models scaled to the extraction area. 
The temperature, metallicity, and the normalization were allowed to vary freely, as well as the absorption column density.
We checked the deprojected temperature within the central cool core region ($<7$ kpc) from the inner three annuli. 
We extracted one spectrum from the region outside the cool core (from $7-180$ kpc) for deprojection purpose. We then obtained the deprojected temperature
using a model-independent approach \citep[e.g.,][]{Dasadia16, Liu_3C88}, which can account for point sources and chip gaps, by fitting
the spectrum with an absorbed thermal component and accounting for the emission contribution from the outer spherical shell.
Fig. \ref{Tprofile} shows the azimuthally-averaged radial profiles of the projected temperature and abundance, as well as the deprojected temperature
and abundance in the inner $\sim7$ kpc. Since the deprojected abundance cannot be well constrained, 
we fixed it at the value close to that from projection analysis.
With the new deep \chandra\ observation, the temperature profile in the central $\sim11$ kpc region can be resolved 
into 4 data bins. We found the temperature in the innermost bin is $1.17^{+0.15}_{-0.14}$ keV. The temperature then rises
outwards to $4.65^{+0.57}_{-0.46}$ keV at $\sim9$ kpc and remains relatively constant towards $\sim180$ kpc. The abundance profile
outside $\sim7$ kpc follows the general trend found in most of galaxy clusters and it decreases 
from $1.23^{+0.43}_{-0.35}$ Solar abundance to $0.35^{+0.12}_{-0.11}$ Solar.
However, in the central $\sim7$ kpc, the abundance is relatively flat in the central two innermost bins, then rises to
a high value of $\sim3.0$ Solar, though with a large error bar. The temperature and abundance profiles suggest that 3C~129.1 is a
corona system. The elevated abundance at $\sim5$ kpc could be due to that the high-abundance gas in the inner region has been pushed outward.

\begin{figure*}
\begin{center}
\vspace{-1cm}
\hbox{\hspace{-5px}
\includegraphics[scale=0.44]{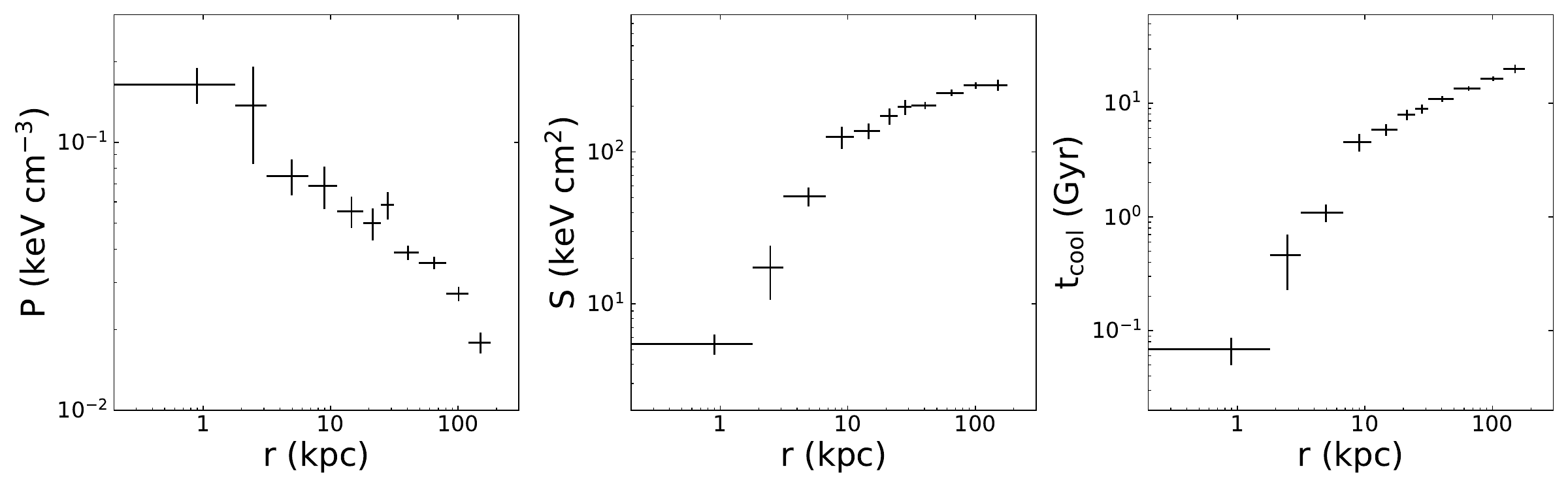}
}
\vspace{-10px}
\caption{
	Radial profiles of pressure (left), entropy (middle), and cooling time (right) of the cluster 3C~129.1, after deprojection.
}
\label{tcool}
\end{center}
\vspace{-0.2cm}
\end{figure*}

Fig. \ref{tcool} shows the profiles of pressure, entropy, and the cooling time for the cluster 3C~129.1.
As mentioned in section \ref{size}, we obtained the gas density by deprojecting the surface brightness profile.
We calculated the pressure as $P=nkT$, the entropy as $S=kT/n_{e}^{-2/3}$, and the cooling time as 
$t_{cool}=\frac{3nkT}{2n_{e}n_{H}\Lambda(T,Z)}$, where $n$ is the total number density, $n_{e}$ and $n_{H}$ are the electron
and proton densities, $\Lambda(T,Z)$ is the cooling function for a given temperature and metal abundance.
The entropy in the innermost bin is $\sim5.5$ keV cm$^{2}$. Starting from $\sim2.5$ kpc (the second data bin) the entropy
increases suddenly from less than 20 keV cm$^{2}$ to $\sim126$ keV cm$^{2}$ at $\sim9$ kpc (the fourth data bin). 
If we fit the entropy in the innermost two data bins and the outermost four data bins with a power-law, 
the entropy in the middle region clearly shows enhancements 
(e.g., $\sim70-90$ keV~cm$^{2}$ in each data bin within $\sim7-30$ kpc), 
suggesting the gas has been heated. The central cooling time is less than 1 Gyr.
The cooling radius, defined as the radius where the cooling time is shorter than 4 Gyr, is estimated to be $8.3\pm1.1$ kpc.
We derive an average temperature of $1.6^{+0.24}_{-0.25}$ keV and a bolometric luminosity  
of $2.9\times10^{41}$ erg~s$^{-1}$ within the cooling radius. 
Assuming the cavity has an ellipsoidal geometry with the projected axis length of $\sim2.9$ and $\sim1.8$ kpc
and a distance of $\sim6.3$ kpc from the nucleus, we estimated a total enthalpy
($H = 4PV$, where $P$ is the azimuthally-averaged pressure at the radius of the centre of the cavity) of $\sim7\times10^{56}$ erg.
The buoyant rise time of the bubble is calculated at its terminal velocity $\sim(2gV/SC)^{1/2}$, where $V$ is the volume of the bubble,
$S$ is the cross section, $C=0.75$ is the drag coefficient, and $g$ is the gravitational acceleration \citep{Churazov2001}.
Assuming hydrostatic equilibrium and spherical symmetry, we estimated the gravitational acceleration as $g=d\Phi/dr=-\frac{1}{\rho}\frac{dP}{dr}$,
where $\Phi$ is the gravitational potential, $\rho$ is the density, $P$ is the pressure.
We obtained a buoyant rise time of $\sim10^{7}$ yr and a cavity power of $\sim2\times10^{42}$ erg~s$^{-1}$.
Based on the $P_{\rm radio}-P_{\rm cav}$ relation \citep[e.g.,][]{Cavagnolo2010}, the jet power is estimated to be $\sim10^{44}$ erg s$^{-1}$,
which is much larger than the observed cavity power. This can be due to a few factors.
First, any potential cavities at faint regions away from the bright center can be difficult to be detected with the limited sensitivity of current data. Indeed, the radio lobes at low frequencies are much bigger than the detected X-ray cavities.
Second, the actual cavity may be larger than the one assumed in the calculation due to the projection effect.
Third, old cavities may have already dissipated and mixed with surroundings.

\section{Discussion}
\label{Discussion}
\subsection{Transition of Cluster Cool Cores}
\begin{figure*}
\begin{center}
\hbox{\hspace{5px}
	\vspace{-0px}
\includegraphics[scale=0.52]{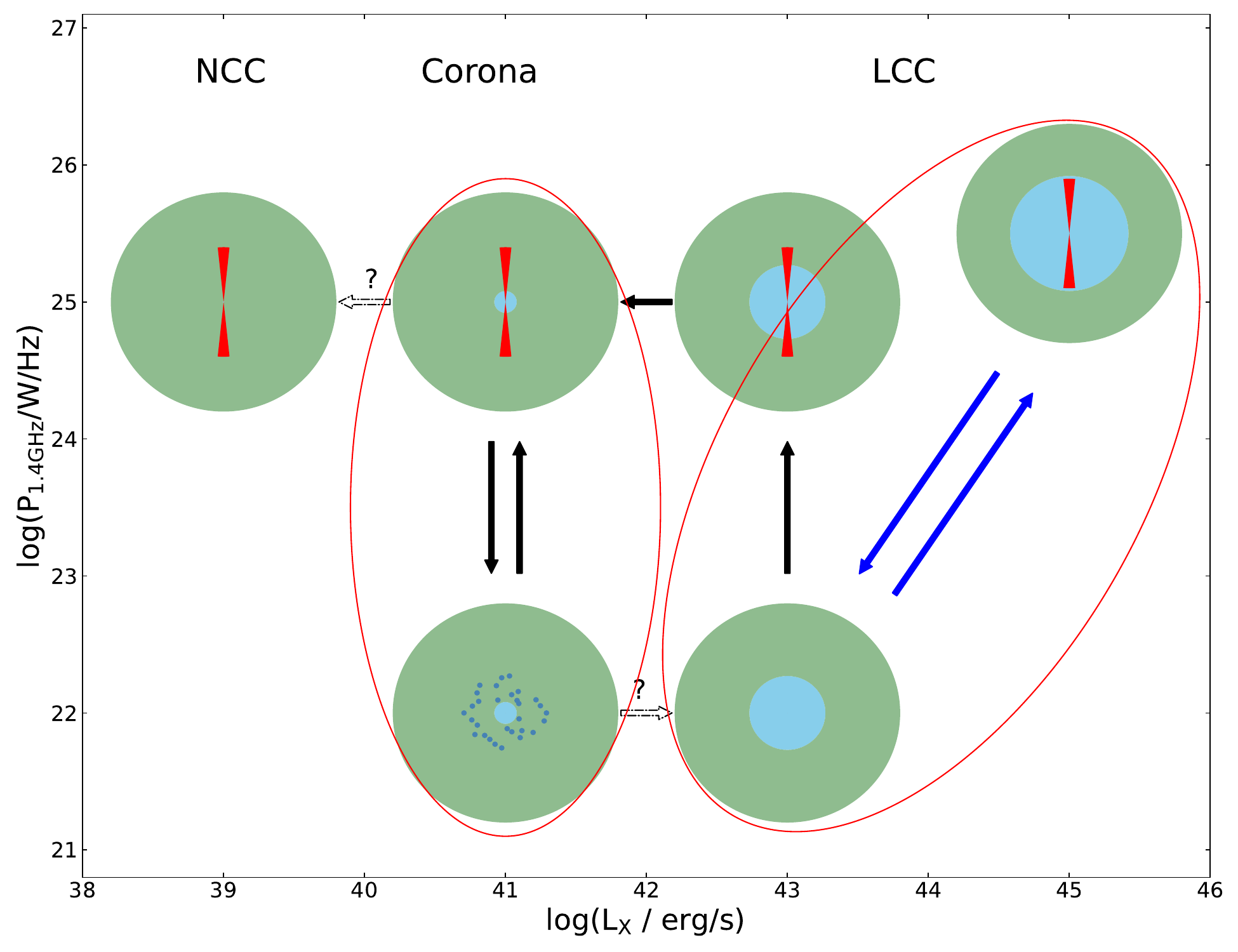}
}
	\caption{Cartoon plot showing the possible transitions of cluster CCs on the $L_{X}-P_{radio}$ plane
	due to radio AGN feedback in galaxy clusters.
	Two red ellipses represent corona clusters and large CC clusters. 
	Large CC clusters are subject to self-regulated radio AGN feedback. The radio jets can switch on and off in one AGN 
	feedback loop. This self-regulated AGN feedback loop in large CC clusters is shown as blue arrows on the right.
        Occasionally, radio quiescent, large CC clusters can have a strong AGN outburst, 
	during which the powerful radio jets penetrate through the inner CC, and deposit a large amount of energy into the outer regions. 
	The CC can be disrupted and turned into a CC remnant or corona (If the CC is totally destroyed, an NCC will be formed).
        The corona can serve as a mini CC and provide fuel for the central SMBH. Due to the relatively small amount of gas, the central 
	radio AGN fades quickly and becomes radio quiescent. Likely aided by residual cool clumps 
	surviving from the previous heating event, the hot gas outside the central corona can develop local thermal instabilities 
	and cool down, represented by the small blue dots around the central corona. If the gas in the corona continues to feed the central SMBH, 
	the radio AGN may switch on again. Or the corona may cool down, reassemble with the cold clumps, which fall back from outside, and develop into a large CC again. The proposed transitions of large CCs and coronae are shown in the solid black arrows. Empty arrows with question marks show a possible transition.
}
\label{cartoon}
\end{center}
\vspace{-10px}
\end{figure*}

Many observational and theoretical studies have suggested that accretion of cold gas fuels AGN feedback in clusters of galaxies 
\citep[e.g.,][]{Pizzolato2005,Gaspari12a,Gaspari2013,Li2014a,Voit15b,McNamara2016,Yang2016a,Wang2019,Liu_3C88}.  
The accretion rate of the central SMBH can be boosted by orders of magnitude via a process known as 
CCA/cold gas precipitation, and a large amount of energy can be released through radio jets.
In this section, we discuss the possible transitions of cluster CCs due to radio AGN feedback.
For illustration, a cartoon plot of cluster CC transitions is shown in Fig.~\ref{cartoon}.
In this study we classify the clusters into two categories, i.e., clusters with the large CC and clusters with
the corona, as shown with two ellipses in Fig.~\ref{cartoon}.

Large CC clusters are in general subject to self-regulated radio AGN feedback. 
Starting with a large CC cluster with its central SMBH dormant, the hot gas in the X-ray CC 
cools via top-down multiphase condensation, thus providing the fueling for the central SMBH through the CCA rain.
Once the SMBH ignites and becomes active, it 
releases a large amount of energy through radio jets that can heat the ICM. Radiative cooling can be suppressed or quenched.  
The heating process will last until the fuel becomes depleted and the SMBH becomes quiescent again. 
This standard maintenance mode AGN feedback self-regulated loop in galaxy clusters \citep[e.g.,][]{Gaspari17} is illustrated with
blue arrows in Fig. \ref{cartoon}.
During this feedback cycle, the radio luminosity of central AGN can increase and decrease, depending on whether the
central SMBH is in an active or quiescent state. 
The CCs can shrink or grow depending on if it is heated or not by the energy released from the central AGN. 
For clusters with very large CCs, e.g., A478, A2029, and the Perseus cluster,
the central gas pressure and density are very high. It is difficult for the radio jets to penetrate through the 
high-pressure CCs and deposit a large portion of energy into the outer part of the CCs.
We do not expect these very large CCs will change by orders of magnitude.

Although corona systems are different from large CC systems in their CC sizes and luminosities,
we take them as different stages of CCs and propose a possible transition scenario for them.
Similarly, starting with a large CC cluster with a radio quiet AGN, occasionally, the cluster can
have a very powerful AGN outburst \citep[e.g.,][]{Li2017}. If the CC is not too large, the energy released can destroy a large portion of 
the CC, as long as the radio jet can penetrate through the central region and deposit most of its energy outside, 
e.g., heat and evacuate the cool gas outside the central 10 kpc.
In fact, this can be achieved for a radio AGN as long as a) the radio jet is very collimated with a small opening angle; b) there is
no relative motion between the jet and ambient gas; c) the jet is fast with a large Mach number so that the
jet can penetrate the central ISM quickly and deposit most of its kinetic energy at large radii 
\citep[e.g.,][]{Soker2016,Yang2019}.
In this way the central region of the CC can be kept intact and the corona-like cool cores can be formed.
(If the central region of the CC is also heated and disrupted, an NCC cluster may be formed as shown with the empty arrow on the
left in Fig. \ref{cartoon}.)
An example with the similar process is the Cygnus~A cluster, whose host galaxy is known as the nearest powerful FR II radio galaxy.
The radio jets, cocoon shock, X-ray jets and cavities, and the central source with very bright X-ray emission show the activity of AGN feedback
\citep[e.g.,][]{Young2002,Smith2002,Wilson2006,Snios2018}. The radio jets penetrate through the central region with a total projected extend of $\sim130$ kpc
from west to east. The thermal gas lying within the elliptical cocoon shock exhibits a rib-like structure, which is the debris resulting from the disintegration of
the cool core by the radio jets \citep{Duffy2018}. While for the thermal gas close to the nucleus 
(3\arcsec\ to 6\arcsec, $\sim3.3-6.5$ kpc from the nucleus), we obtained a temperature of $\sim1.6$ keV 
by masking out the central 3\arcsec\ region due to the very bright central source, and using a local background from 6\arcsec\ to 9\arcsec\ region. For the analysis,
we followed the standard \chandra\ data reduction and re-analysed one \chandra\ observation of Cygnus~A with ObsID 17512. We fitted
the spectrum with an absorbed thermal component plus a power-law component.
Although Cygnus~A is a large cool core cluster in our sample, it represents an example
where the radio jets are heating beyond the central region where there still exists low entropy gas.

The corona of the central BCG can serve as a mini-cool core and continue to provide fuel to the central SMBH. 
Note that in this process, most of the energy released by central radio AGN should be deposited outside the corona,
otherwise, the corona can be destroyed, since the jet power greatly exceeds its cooling luminosity as shown in Fig. \ref{LxPr}.
However, the corona cannot support radio AGN for very long, as indicated by the fraction of radio strong
corona systems in section \ref{size}.
When the central engine runs out of fuel, the SMBH will shut down and the radio luminosity will eventually drop.
Radio strong corona clusters will evolve into radio weak corona systems. 

Finally, the central region of the corona also cools down (as discussed in Section \ref{SNheating}) and provides fuel to the
central SMBH. Once the SMBH ignites, the radio AGN will switch on again.
Moreover, the X-ray cool core remnants from the previous heating episodes may have residual cool clumps that survived, without raising the X-ray surface brightness much. Without the heating from the central AGN, cooling outside the corona can proceed, aided by the possible residual cool clumps and local thermal instability.
Eventually, if the corona can reassemble with the cool gas that falls back from outside, a large CC may be formed again. 
The possible transitions of large CC and corona systems are shown in the black arrows 
in Fig. \ref{cartoon}.

Simulations have shown that large CCs can be destroyed by high-energy, head-on major mergers and CC clusters
can be transformed into NCC clusters \citep[e.g.,][]{Hahn2017}.
Using binary merging cluster simulations, \citet{Valdarnini2021} found that CCs may be resilient to off-axis mergers with low mass ratios,
and CC can survive depending on the initial mass ratios and the angular momentum of the system, as well as in other simulations.
One of the important features from a nonzero impact cluster mergers is the gas sloshing \citep[e.g.,][]{Markevitch2007}, which has been proposed as another heating source \citep[e.g.,][]{ZuHone2010}. Our results show that most of clusters with large
offsets have small cool cores, which suggests that gas sloshing could be an effective way to reduce the size of CCs.
For example, the recent X-ray studies of galaxy cluster A119 (cluster in our sample with large offset and small CC) 
found two cold fronts, possibly associated with large-scale gas sloshing \citep{Watson2023}.
In addition, the reduction of CC size may be the joint effect of gas sloshing and AGN feedback \citep[e.g.,][]{Rasia2015}.
The large-scale gas-sloshing caused by cluster passage can cause a disturbance in the ICM and heat cluster CC. If the central SMBH is 
triggered during this process, the energy released from the AGN can be deposited in the ICM and the CC will be further disrupted.

\subsection{Duty Cycle}
\label{duty}
In our study, we define the duty cycle of radio AGN as the fraction of the time when the galaxy hosts a radio-loud AGN,
which is crucial to quantify the importance of the AGN feedback.
Previous studies show that the duty cycle of radio galaxies based on their radio power has a dependence on the stellar mass of 
the host galaxies \citep[e.g.,][]{Best05,Best07,Shabala2008}.
In our sample, we found that the fraction of BCGs with radio-loud AGN in the relaxed, small CC systems is 
$\sim29$\% ($P_{\rm 1.4 GHz} >10^{23}$ W Hz$^{-1}$). 
The value is consistent with the value for BCGs with a stellar mass of $\sim10^{11.5}$ M$_{\odot}$ \citep[e.g.,][]{Best07},
with the same radio power threshold for radio-loud AGN.
For the relaxed, large CC clusters, the fraction of BCGs with radio-loud AGN is much larger, with a value of 
$\sim71$\% ($P_{\rm 1.4 GHz} >10^{23}$ W Hz$^{-1}$).
This is consistent with the findings of previous studies that clusters with strong CCs tend to host a central radio luminous
source \citep[e.g.,][]{Burns90,Mittal2009}.
Recently, using the \gmrt, \citet{Kolokythas2018} studied a complete sample of local groups 
with high-richness and found a high radio-detection rate of 92\%. 
\citet{Grossova2022} presented Karl G. Jansky Very Large Array radio observations
of a sample of the 42 nearest optically and X-ray brightest early-type galaxies, and found 41 of 42 galaxies have radio detections.
But if we take the same radio power threshold (i.e., $P_{\rm 1.4 GHz} >10^{23}$ W Hz$^{-1}$), the fractions of galaxies that are
radio-loud in these two studies decreases to $\sim11$\% and $\sim26$\%, respectively.
Moreover, using a complete sample of clusters, \citet{Birzan2012} found that the duty cycle for radio-mode feedback
is at least $63$\%, which is slightly smaller than the fraction of BCGs with radio-loud AGN
(71\% for $P_{\rm 1.4 GHz} >10^{23}$ W Hz$^{-1}$) in our large CC clusters.
However, we should note that the duty cycle for radio-mode feedback used in their study is the fraction of the time
when a cluster possesses bubbles inflated by the central radio source.

In addition, during the life-cycle of a radio AGN, it experiences different phases, e.g., from a young, newly born radio source
to an evolved radio source, then to the remnant phase when AGN activity stops, and to a restarted phase 
when the central AGN is active again.
The radio morphology and spectra contain essential information about the radio sources and can be used to identify
which phase the radio source is currently in.
Using a method combining spectral ageing and dynamical models, \citet{Turner2018} studied the galaxy group B2 0924+30 
with the remnant and restarted radio-loud AGN to constrain its duty cycle, which is defined as $t_{on}/(t_{on}+t_{off})$, 
where on and off represent the radio jet being switched on and off.  
Note that the paper did not give a quantitative radio power threshold corresponding to the active and quiescent stages, because
in the spectral ageing model including radio source evolution, the radio luminosity is a function of both the 
age of radio source and jet power, and a low-powered source can maintain its luminosity long after the jet is switched off.
In their study, the remnant AGN is found to have an active age of 50 Myr and a total age of 78 Myr, corresponding to a duty cycle of $\sim64$\%.
However, if considering a previous AGN outburst in the model, they found that the time between two outbursts is at least 330 Myr.
Therefore, the duty cycle of remnant B2 0924+30 is less than $15\%$.
\citet{Biava2021} used new LOFAR observations at 144 MHz together with archival radio data at higher frequencies to investigate
the spectral properties of galaxy cluster MS~0735.6+7421, which has one of the most powerful known AGN and two pairs
of X-ray cavities \citep[e.g.,][]{Vantyghem2014}. 
Similarly, by fitting the lobe spectra but using a single particle injection model, they derived the spectra ages for outer and intermediate lobes,
which represent different jet episodes.
They found that the two episodes of jet activity are separated by a brief quiescent phase and the duty cycle is close to unity.

Based on the fractions of BCGs with radio-loud AGN, the duty cycle for small CC clusters is 
less than half of that for large CC clusters.
This suggests that the radio activity of BCGs is affected by the properties of the surrounding gas beyond the central $\sim10$ kpc. 
Strong radio AGNs in small X-ray CCs fade more rapidly than those embedded in large X-ray CCs.
However, it is not very clear what causes this difference in duty cycle in the small and large CC systems, 
and if the difference is related to the available amount of cold gas serving as the fuel for the central SMBH.
A detailed study of the inventory of cold/warm gas in corona systems will provide essential information to answer this question.

\subsection{Heating in the Cool Core Coronae}
\label{SNheating}
Another interesting question is how a corona system can maintain its steady state.
As shown in Fig. \ref{LxPr}, the cavity power estimated from the radio luminosity for corona systems is much larger
than the cooling luminosity. Therefore, most of the energy released from radio AGN must be deposited into
the ICM outside the central corona, which can be achieved in certain conditions \citep[e.g.,][]{Soker2016}.
In addition, the total heat flux across the surface of a cool corona surrounded by a hot ICM can be estimated
as $8.2\times10^{42}\ (T/5\ \textrm{keV})^{3/2}(n_{\textrm{ICM}}/10^{-3}\ \textrm{cm}^{-3})(r/3 \textrm{kpc})^{2}$ ergs~s$^{-1}$ based
on the saturated conductive heat flux in \citet{Cowie1977}, which is much larger than
the X-ray luminosity of corona (e.g., see table \ref{tab3}) within a typical hot cluster.
Therefore, the thermal conductivity is required to be significantly reduced for the survival of 
cool core coronae \citep[e.g.,][]{Vikhlinin2001, Sun2007}.
In this way, the central corona can serve as a steady small cooling flow and fuel the central SMBH without disruption. 
However, if the energy released from the central radio AGN is used to heat the IGM or ICM at large radii, there should
be another heating source to balance the cooling.
\citet{Voit2020} proposed a black-hole feedback valve mechanism, in which SNe heating in galaxies with
a large central velocity dispersion can lead to a quasi-steady state with a modest cooling flow at small radii, while
AGN feedback appears to regulate the CGM pressure or heat the ICM \citep[e.g., see review in][]{Donahue2022}.

Here we consider the requirements on SNe feedback, such as the heating from explosions of white-dwarf SNe (SNIa) to balance the cooling
within the central coronae. 
\citet{Sun2007} found that the SNe inside the coronae generally have enough energy to balance cooling, although the required
fraction of the SNe energy required to be deposited into the coronal gas increases with the galaxy's luminosity.
We adopted the same SNe rate in early-type galaxies \citep{Cappellaro1999}, $0.166 h^{2}_{70}$ per 100 yr per 10$^{10}$ $L_{B\odot}$, and assumed a kinetic energy of 10$^{51}$ erg per SN. For a stellar population of age $\sim10$ Gyr and 
a stellar mass of $\sim10^{11.5}$ $M_{\odot}$, the SNIa kinetic power is $\sim2\times10^{41}$ erg s$^{-1}$, 
assuming a stellar mass-to-light ratio of $10$ in the B-band for early-type galaxies \citep[e.g.,][]{Faber1979, Nagino2009}.
A recent study showed that the observed SNIa rate for a stellar population among elliptical galaxies in galaxy clusters is 
$\sim3\times10^{-14}(t/10 Gyr)^{-1.3}$ yr$^{-1}$ $M_{\odot}^{-1}$ \citep{Friedmann2018}. 
Using the updated SNIa rate, we obtained the SNIa feedback power of $\sim3\times10^{41}$ erg s$^{-1}$.
Since the dividing X-ray bolometric cooling luminosity for large CCs and coronae
is $\sim 10^{42}$ erg s$^{-1}$ (the conversion factor from $0.5-2$ keV luminosity to bolometric luminosity for low-temperature corona gas is $\sim2$), 
the SNIa heating is enough to balance the cooling for many corona systems, as long as the energy of SNIa can be efficiently coupled into the corona gas.
However, the SNe heating should follow the stellar light profile, which is shallower than the X-ray emission profile
of coronae \citep[e.g.,][]{Sun2007}.
Thus, it can be challenging to match SN heating and cooling spatially, so a corona may still cool around the nucleus and provide fuel for the central SMBH.

\section{Conclusion}
\label{Conclusion}
In this paper, we used a large sample of 108 nearby clusters ($z<0.1$) with $kT>3$ keV to study the effect of
AGN heating and the transitions of cluster CCs. The main conclusions are the following.
\begin{itemize}
	\item For galaxy clusters with small offsets ($<50$ kpc) between the central BCG and the X-ray cluster centre, about 40\% of them have small CCs. While for clusters with large offsets, most of them (14 out of 17) have small CCs. This suggests that cluster mergers or sloshing are efficient in reducing the CC size.
	\item Using the CC size or CC luminosity to classify small and large CCs gives consistent results.
		Large CC clusters generally show a weak correlation between the CC luminosity and the radio luminosity.
		The inferred kinetic power, estimated on the basis of the $P_{radio}-P_{cavity}$ relation, generally exceeds the CC luminosity. Small CC clusters do not show such a correlation, and the inferred kinetic power could greatly exceed the CC luminosity.
	\item In clusters with large offsets, the fraction of radio strong systems ($P_{1.4 \rm GHz}>10^{23}$ W Hz$^{-1}$) is $\sim53$\%.
		In clusters with small offsets, large CC systems show a relatively continuous distribution of radio power, while small CC systems are mostly 
		radio weak (71\% for $P_{1.4 \rm GHz}<10^{23}$ W Hz$^{-1}$).
	Furthermore, there is a lack of small CC clusters with intermediate radio power of $2\times10^{23}-2\times10^{24}$ W Hz$^{-1}$. 
		This suggests different radio AGN feedback processes in large and small CC clusters. The radio AGN stage is more continuous in large CCs, while in small CCs, the radio AGNs spend a much shorter time in radio strong stage.
	\item We estimated the circumnuclear X-ray luminosities, within 1 kpc of the BCGs, and found that there are no systems with low radio power ($P_{1.4 \rm GHz}<10^{24}$ W Hz$^{-1}$) while having a strong circumnuclear X-ray luminosity ($L_{X}>10^{41}$ erg s$^{-1}$). 
	\item We presented a detailed study of the galaxy cluster 3C~129.1. The small central cooling time, low central entropy, abundance bump, and elevated entropy profile outside the central core ($>\sim9$ kpc) suggest that it
		is a CC remnant, possibly disrupted by a radio AGN outburst.
	\item We outlined the evolutionary transitions that cluster cores may undergo in their thermal states.
		Clusters with luminous CCs are subject to a self-regulated feedback cycle. If, occasionally, the radio AGN outburst is very powerful,
the collimated radio jets can penetrate the central part of CC and deposit a large amount of energy outside \citep[e.g.,][]{Soker2016,Yang2019}. In this way, a large portion of the CC can be disrupted and a corona class object can be formed with the gas outside the central $\sim10$ kpc being heated and evacuated. However, the corona cannot support radio AGN for very long, and the radio luminosity of the central BCG fades quickly once the SMBH is out of fuel. 
		If the hot gas in the corona can cool and fuel the central SMBH, its radio activity can be enhanced again. Outside of the central corona, if there are residual cool clumps surviving from the previous heating episode, they may grow by cooling and may also rain down towards the central corona to increase the cool core size.
	\item Based on the fraction of BCGs with radio strong AGN, we found that the
duty cycle in relaxed, small CC clusters is $\sim29$\%
		($P_{\rm 1.4GHz}>10^{23}$ W Hz$^{-1}$), while the duty cycle for large CC clusters is $\sim71$\%, 
		respectively.
		The low duty cycle in small CC clusters suggests that the radio activity of BCGs is affected by the properties of the
		surrounding gas beyond the central $\sim10$ kpc, and strong radio AGN in small X-ray CCs fade more rapidly than
		those embedded in large X-ray CCs.
		However, the origin of the different duty cycles between small CC clusters and large CC clusters is unclear. Detailed studies on the inventory of cold/warm gas in corona systems will help to address this question.
\end{itemize}
This study used data carefully collected from the \chandra\ archive, but it is still limited by the detection ability of current X-ray instruments,
e.g., we have to set a redshift limit of 0.1 in order to resolve the gas properties in the central region of galaxy cluster. Future X-ray missions with larger collecting
power and better spatial resolution will help to observe more corona systems at low and higher redshift and to study their properties and cool core transitions.

\section{Acknowledgements}
We are grateful to the referee for comments that improved this manuscript. 
This work is supported by the National Science Foundation of China (12233005 and 12073078), the science research
grants from the China Manned Space Project with NO. CMS-CSST-2021-A02, CMS-CSST-2021-A04 and CMS-CSST-2021-A07. 
WL is supported by Jiangsu Provincial Double-Innovation Doctor Program (JSSCBS20211400). 
MG acknowledges partial support by HST GO-15890.020/023-A and the \textit{BlackHoleWeather} program.
CS was supported in part by the {\it Chandra} grant GO1-22120X.
The NRAO VLA Archive Survey image was produced as part of the
NRAO VLA Archive Survey, (c) AUI/NRAO. The National
Radio Astronomy Observatory is a facility of the National
Science Foundation operated under cooperative agreement
by Associated Universities, Inc. We thank the staff of the
GMRT that made these observations possible. GMRT is run
by the National Centre for Radio Astrophysics of the Tata
Institute of Fundamental Research.

\section*{Data Availability}
The Chandra raw data used in this paper are available to download at the HEASARC Data Archive website (https://heasarc.gsfc.nasa.gov/docs/archive.html).
The reduced data underlying this paper will be shared on reasonable requests to the corresponding authors.
This research has made use of software provided by the Chandra X-ray Center (CXC) in the application package CIAO.

\bibliographystyle{mnras}
\bibliography{my}

\appendix{}
\section[]{}
\label{appendix}
\begin{figure}
\hbox{\hspace{-5px}
\includegraphics[scale=0.53]{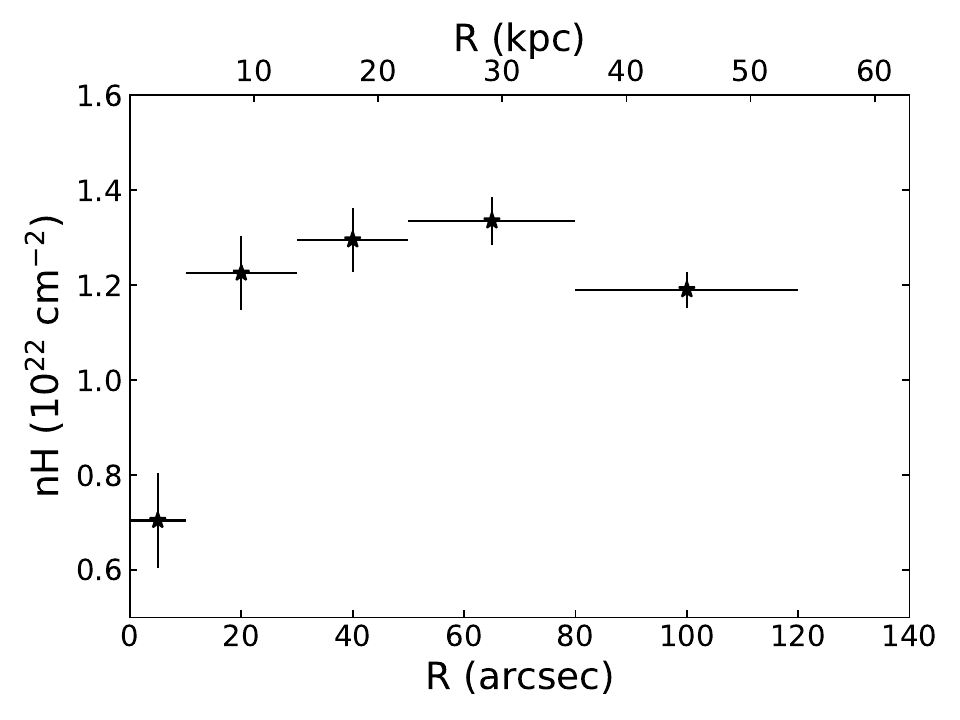}
}
\caption{
	The fitted absorption column density as a function of radius for the 3C~129.1 cluster. 
}
\label{nh}
\end{figure}

Since the cluster 3C~129.1 lies close to the Galactic plane, the Galactic absorption is high and its accurate value is essential for our spectral analysis.
We examined the X-ray absorption column density towards 3C~129.1. 
The weighted column density of the total Galactic hydrogen, including both the atomic hydrogen and molecular hydrogen,
is $7.36\times10^{21}$ cm$^{-2}$, calculated using the ``NHtot'' tool\footnote{http://www.swift.ac.uk/analysis/nhtot/index.php}
\citep{Willingale13}. 
We then checked the nH profile by fitting the spectra extracted from several annular regions centred on the cluster 3C~129.1
from two new ACIS-S observations.
The model includes an absorbed thermal component for the diffuse emission from the cluster plus the diffuse X-ray 
background component. The background was modeled and re-scaled based on the local X-ray background estimate. 
The absorption column density, temperature, abundance, and normalization of the cluster emission were allowed to vary. 
We plotted the best-fitting values of the absorption column density in Fig. \ref{nh}. 
The absorption column density in the central 10 arcsec is consistent with the weighted column
density of the total Galactic hydrogen. It then rises quickly to a high value of $\sim1.2\times10^{22}$ cm$^{-2}$ and becomes
relatively flat outside the central 10 arcsec.

In our analysis, we tried both fixing the absorption column density at the weighted value of $7.36\times10^{21}$ cm$^{-2}$,
and letting it vary freely. We found that freeing the column density in the spectral fitting gives a much lower C-statistic. 
Therefore, in the text we show the result obtained from spectral fitting with free absorption column densities for 3C~129.1.

\end{document}